\documentclass[12pt]{article}

\usepackage{graphicx}

\begin{document}

\begin{center}

\Large{\bf On the Role of Caustic in Solar Gravitational Lens Imaging}

\vspace{1cm}

Igor Loutsenko 

\vspace{5mm}

{\small Laboratoire de Physique Math\'ematique,\\
CRM, Universit\'e de Montr\'eal\\
e-mail: loutseni@crm.umontreal.ca\\[1mm]}

\vspace{1cm}

\begin{abstract}

{

We consider scattering of electromagnetic waves from a distant point source by the gravitational field of the sun, taking the field oblateness due to the quadrupole moment of the sun into account. Effects of the field oblateness can play an important role in the high resolution solar gravitational lens imaging in the sub-micrometer wavelength range of the electromagnetic spectrum.
}

\end{abstract}

\end{center}

\vspace{1cm}

\section{ Introduction}

\vspace{5mm}

The idea of using the sun as a powerful telescope goes back to Eshleman \cite{Eshleman}: The gravitational field of the sun acts as a spherical lens to magnify the intensity of radiation from distant objects along a semi-infinite focal line with the nearest point of observations being about 550AU (for a general introduction, see e.g. \cite{L}, \cite{TA}, \cite{M}).
For example, the intensity from a distant point source of electromagnetic (EM) radiation at $\lambda=1\mu{\mathrm m}$ wavelength can be pre-magnified by the sun gravitational lens up to $\mu\approx 10^{11}$ times. Depending on observation device, the resolvable angle between two point sources at this wavelength could be as small as $10^{-10}$ arcsec.

Recently, properties of the solar gravitational lens attracted attention both due to discovery of numerous exo-(and possibly earth-like) planets and the success of the Voyager-1 spacecraft, presently operating at about 140AU. Possibilities of mega-pixel imaging of such planets from the focal line of solar gravitational lens are now being discussed.

In the present work we consider effects of oblateness of the gravitational field and that of rotation of the sun on the image formation and the diffraction pattern of the lens. Although the quadrupole moment of the sun is very small, effects of oblateness, nevertheless, turn out to be important: The focal line caustic unfolds and can have several hundred meters in cross section at the distances up to several thousand AU from the sun. Moreover, for the wavelengths in a micrometer range, the diffraction pattern of the point monochromatic source changes significantly and the maximum of the amplification of the EM energy flux radiated by such a source can decrease up to several orders of magnitude, depending on direction of observation and the distance between the sun and an observer.

We stress that the last statement concerns the maximal magnification of density of flux radiated by a point source, and not the intensity magnification for a realistic extended object. Since the ``focal blur" of the gravitational lens is comparable with the size of the whole image \footnote{Here ``image" means the heliocentric projection of the source, which shows up as a ``high intensity" region. }, there will be no significant difference in the flux magnification for realistic objects. For instance, for visible light from exo-planets of interest, the maximal pre-magnification of the flux by the sun gravitation is about $10^5$-$10^6$  \cite{L}, in difference from $\mu \sim 10^{11}$ for a point source (see e.g. \cite{TT}, \cite{T}), and it is determined mainly by the geometrical optics.  However, effects caused by the field oblateness should be accounted for in de-convolution of images of objects of interest: In the case of exo-planets, such effects can become important already at the hecto-pixel level of resolution of de-convolution (see Section 6).

Lensing by oblate objects was extensively studied in the literature in general and (to a lesser extent) in connection with the solar gravitational lens.  For instance, in \cite{Eshleman} some heuristic considerations of effects caused by the field oblateness were presented. In the work \cite{Koechlin} some estimates also were made that led to a conclusion that the oblateness of the sun  has a negligible effect. This conclusion has been drawn from the computation of difference in deflection angles in the sun equatorial and polar planes, which is based on a heuristic model of the gravitational field of the sun  \footnote{In \cite{Koechlin}, the model of two spheres of half of density offset by the distance comparable with that determined by oblateness is used. This leads to an underestimate in the corrections to deflections by about three orders of magnitude in comparison with e.g. computations made by Epstein and Shapiro \cite{ES}. The latter are based on the model with correctly estimated quadrupole moment. As a consequence, the transverse abberation was also underestimated by about three orders of magnitude in \cite{Koechlin}. The correct expression for the size of the abberation/caustic was derived by e.g. Eshleman et al in \cite{ETF}.}. Rigorous estimates of the deflection corrections due to the quadrupole moment of the sun have been done earlier, for instance, by Epstein and Shapiro \cite{ES}. The correct expression for the size of the abberation/caustic was derived by e.g. Eshleman et al in \cite{ETF}. Our estimates are based on direct computations using methods of the uniform (caustic) expansions in the geometrical theory of diffraction (for detailed introduction to these methods see e.g. \cite{BK}), rather than on empirical approaches.

It is also worthy to note that detailed studies of the wave-optical aspects of the solar gravitational lensing in the sub-micrometer diapason of the EM spectrum have been done mainly for the spherically symmetric case \cite{HS1}, \cite{HS2}, \cite{T}, \cite{TT}.

This work is organized as follows: In the next section we introduce notations and review geometrical optics of a spheroid gravitational lens with small quadrupole moment. In the 3rd section we consider the wave effects, deriving the gravitational point spread function in the form of a one-dimensional integral and making its numerical evaluation. Three limiting cases when this integral can be computed analytically (the cases of observations in directions (1) close to the sun polar axis, (2) close to the sun equatorial plane, and (3) at the central axis of caustic) are considered in the 4th section. A compound system consisting of the gravitational lens and a telescope is considered in the 5th section. A discussion of the role of caustic structure/size in prospective observations as well as suggestions for further studies are made in the concluding section. The main text of the paper is supplemented with two Appendices: In the Appendix 1 we double-check our results with the help of well-known algorithms from the geometrical theory of diffraction and in Appendix 2 we consider effects of refraction in the solar atmosphere (corona).

Concluding this section we would like to mention that the article contains several examples where we estimate our results at the distance 550AU from the sun. Although, due to the brightness of the solar corona, observations will be rather possible at distances $> \approx 1000$AU than at $550$AU, our estimates are scalable: one gets the same estimate for bigger distance and proportionally smaller wavelength.

\section{Geometrical Optics Problem}

We are interested in description of the diffraction pattern of an EM wave scattered by the gravitational field of the sun. For this we first review the geometrical optics counterpart of this problem: namely the deflection of initially parallel light rays coming from an infinitely distant point source.

A trajectory of light in the gravitational field of the sun can be found using post-Newtonian approximations for the null-geodesics of the post-Minkowskii metric element (see e.g.  \cite{SEF}, \cite{LL}, \cite{AK}, \cite{S}, \cite{SMT})
$$
ds^2 = \left(1+2\frac{\Phi}{c^2}\right) (cdt)^2- \left(1-2\frac{\Phi}{c^2}\right) d\vec{r}^2-\frac{8}{c^2}\left(\vec{A}\cdot d\vec{r}\right)dt ,
$$
where $\Phi$ is the scalar (Newtonian) gravitational potential and $\vec{A}$ is the gravitomagnetic vector potential.

In the asymptotically Cartesian heliocentric coordinate system where parallel beams are incoming from $z=-\infty$, the post-Newtonian deflection angle $\vec{\alpha}$, which is the difference between the incoming and outgoing beam direction vectors, equals the two-dimensional gradient of the potential $\Psi$
\begin{equation}
\vec{\alpha}=\nabla \Psi, \quad \nabla =(\partial_x, \partial_y), \quad \Psi=-\frac{2}{c^2}\int_{z_{\rm source}\to-\infty}^{Z} \left(\Phi(x,y,z)-\frac{2}{c} A_z(x,y,z) \right) dz ,
\label{deflection}
\end{equation}
where $Z$ is the $z$-coordinate of an observer. The condition $Z\gg\sqrt{x^2+y^2}\gg r_g$, where $r_g\approx 3\times 10^3$m is the gravitational radius of the sun, is also imposed on eq. (\ref{deflection}).
In this limit the two dimensional gradient of $\Psi$ is independent of $Z$: The potential $\Psi$ is a sum of the $x,y$-dependent ($Z$-independent) and the $Z$-dependent ($x,y$-independent) terms (see below), so it is essentially a two-dimensional potential.

\begin{figure}
  \centering
  \includegraphics[width=150mm]{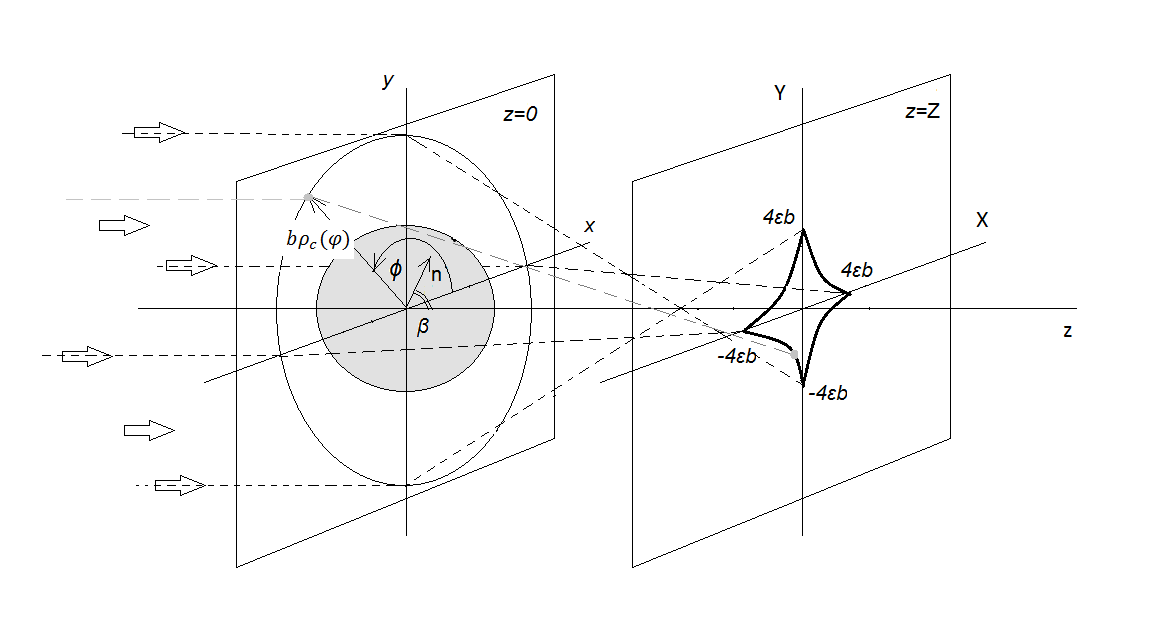}
  \caption{Diagram of the geometrical optics problem. Section of the caustic surface by the observer $z=Z$ plane ($X,Y$-plane) is schematically shown on the right, while the corresponding critical line in the lens $z=0$ plane ($x,y$-plane) is schematically shown on the left (see eqs. (\ref{critical}, \ref{caustic})).}\label{Sun1}
\end{figure}

Here, one can apply the thin lens approximation which leads to the following picture (see Figure \ref{Sun1}): a light ray is incoming from $z=-\infty$ and hitting the $z=0$ ``lens plane" at $(x,y)$. At this plane the ray is deflected by the angle given by the $Z\to\infty$ limit in eq. (\ref{deflection}). Then, it follows that the outgoing ray intersects the observer plane $z=Z>0$ at the point whose position $(X,Y)$ is determined by the extremum of the Fermat potential $S$ (``lens equation"):
\begin{equation}
\partial_x S =0, \quad \partial_y S = 0,
\label{lens1}
\end{equation}
where
\begin{equation}
 \quad S=\frac{(X-x)^2+(Y-y)^2}{2Z}-\Psi(x,y) .
\label{lens2}
\end{equation}
This equation is a manifestation of the Fermat principle for the beam delay time $S/c$.

For a compact lens, a combined contribution to the two-dimensional potential $\Psi$ in eq. (\ref{deflection}) from the dipole terms of $\Phi$ and $\vec{A}$ can be cancelled by a translation. Since the gravitomagnetic field of the sun, produced by its rotation, is a dipole field, without loss of generality we can set $\vec{A}=0$ (for more details see eg \cite{AK}).

The exterior Newtonian potential of the sun can be approximated by that of the quadrupole
\begin{equation}
\Phi(\vec{r})=-\frac{r_gc^2}{2r}\left[1-\frac{I_2}{2}\left(\frac{R_0}{r}\right)^2\left(\frac{3(\vec{n}\cdot\vec{r})^2}{r^2}-1\right)\right] ,
\label{Phi}
\end{equation}
where $\vec{n}$ is a unit vector in the direction of the polar axis of the sun, $R_0 \approx 7 \times 10^8 $m is the sun radius and $I_2\approx 2\times 10^{-7}$ is its dimensionless quadrupole moment . 

Without loss of generality we select the coordinate system where
\begin{equation}
\vec{n}=(0, \sin\beta, \cos\beta)
\label{n}
\end{equation}
with $\beta$ being the angle between the polar axis of the sun and the incoming from $z=-\infty$ beams (see Figure \ref{Sun1}).

Introducing the polar coordinates $(r_\perp, \phi)$ in the lens $z=0$-plane
$$
x=r_\perp\cos\phi, \quad y=r_\perp\sin\phi, \quad r=\sqrt{r_\perp^2+z^2}
$$
and taking (\ref{deflection}), (\ref{Phi}) and the fact that $Z\gg r_\perp$ into account we get \footnote{
For details of this simple computation one can also refer to e.g. \cite{SMT}.} the two-dimensional potential $\Psi$
\begin{equation}
\Psi=2r_g\left(\log\frac{r_\perp}{r_g}-\frac{I_2R_0^2\sin^2\beta}{2r_\perp^2}\cos2\phi\right)+cT(Z) .
\label{Psidimension}
\end{equation}
The term $cT(Z)=r_g\log(-4Zz_{\rm source}/r_g^2)$ can be dropped without loss of generality: It does not affect the geometrical optical values \footnote{The above term can be also dropped in the wave optical computations of the field intensity, which is the square of absolute value of the complex field amplitude, since it contributes only to the common phase factor to the amplitude (see next section).} since it is independent of $x,y$.

In order not to carry numerous constants through the computations, we re-scale both the lens plane and the observer plane lengths with the scaling length parameter \footnote{We would like to stress that $b$ is not an impact parameter. The latter will be introduced further.} $b$
\begin{equation}
b\equiv\sqrt{2r_gZ}.
\label{parameters1}
\end{equation}
The new dimensionless polar coordinates $(\rho, \phi)$ in the lens plane and the dimensionless Cartesian coordinates $(\xi, \eta)$ in the observer plane then read
\begin{equation}
r_\perp =b \rho, \quad (x,y) = (b\rho\cos\phi, b\rho\sin\phi),\quad (X,Y)=(b\xi,b\eta) .
\label{r}
\end{equation}
In these coordinates
\begin{equation}
\Psi = 2r_g\psi, \quad \psi=\log(\rho)-\frac{ \epsilon }{\rho^2} \cos2\phi ,
\label{Psi}
\end{equation}
where
\begin{equation}
\epsilon  = \frac{I_2R_0^2\sin^2\beta}{4 r_gZ}=\frac{I_2R_0^2\sin^2\beta}{2b^2} .
\label{epsilon}
\end{equation}
In the case of the sun $\epsilon\le$ about $10^{-7}$. The small parameter $\epsilon$ is maximal when $\beta=\pi/2$, i.e. when the source is placed in the sun equatorial plane. It decreases as the source is displaced towards the sun polar axis, on which it vanishes (at $\beta=0$): The light radiated by a source from the polar axis is deflected as if the sun were spherically symmetric. So, we will refer to the both situations of $I_2=0$ and $\beta=0$ as the ``spherically symmetric", ``degenerate" or the ``monopole" case. The parameter $\epsilon$ also decreases when the observer plane moves away from the sun (i.e. as $Z$ increases).

It is worth mentioning that one can also account for the light refraction in the solar plasma by adding corresponding correction term to the potential $\psi$. However, this contribution can be discarded for the sub-micrometer diapason of wavelengths. Evaluation of this contribution is given in the Appendix 2.

It follows from (\ref{lens1},\ref{lens2}, \ref{r}, \ref{Psi}) that coordinates ($\rho, \phi$) of the images of the point $(\xi,\eta)$ are solutions of the lens equation, which has the following form in the complex notations
\begin{equation}
\xi+i\eta = \left(\rho-\frac{1}{\rho}\right)e^{i\phi}-\frac{2 \epsilon }{\rho^3}e^{3i\phi} .
\label{xy}
\end{equation}
We recall that the solution of this equation gives the ``impact parameter" $b\rho(\xi, \eta; \epsilon)$ and the corresponding polar angle $\phi(\xi, \eta; \epsilon)$ in the lens plane for the ray(s) arriving to the observer plane at $X=b\xi, Y=b\eta$. Both scaling factor $b$ and a small parameter $\epsilon$ depend on distance $Z$ between the planes.

In the geometrical optics, the inverse intensity magnification equals the ratio of the corresponding surface elements on the observer and the lens planes  (the Jacobian of transformation (\ref{xy}) or the Hessian of the Fermat potential $S$)
\begin{equation}
\mu^{-1}=\left|\frac{\partial(X,Y)}{\partial(x,y)}\right|=\frac{1}{\rho}\left|\frac{\partial(\xi,\eta)}{\partial(\rho,\phi)}\right|=1-\frac{1}{\rho^4}-\frac{12 \epsilon \cos 2\phi}{\rho^6}+{\cal O}(\epsilon^2) .
\label{mu}
\end{equation}
The magnification $\mu$ diverges at the critical line (see Figure \ref{Sun1})
\begin{equation}
\rho=\rho_c(\phi)=1+3 \epsilon \cos 2\phi+{\cal O}(\epsilon^2)
\label{critical}
\end{equation}
or, according to (\ref{xy}), at the astroid (tetracuspid) caustic \footnote{According to the recent data the dimensionless octopole moment of the sun $\sim 10^{-9}$.
Since $\epsilon$ is at most $\approx 10^{-7}$, the caustic cross-section can be considered as a pure astroid for our purposes.} in the observer plane
\begin{equation}
(\xi,\eta)=(\xi_c(\phi),\eta_c(\phi))=4 \epsilon (\cos^3\phi, - \sin^3\phi) .
\label{caustic}
\end{equation}
Here we recall the standard procedure of solving the lens equation (\ref{xy}).

\begin{figure}[t]
  \centering
  \includegraphics[width=175mm]{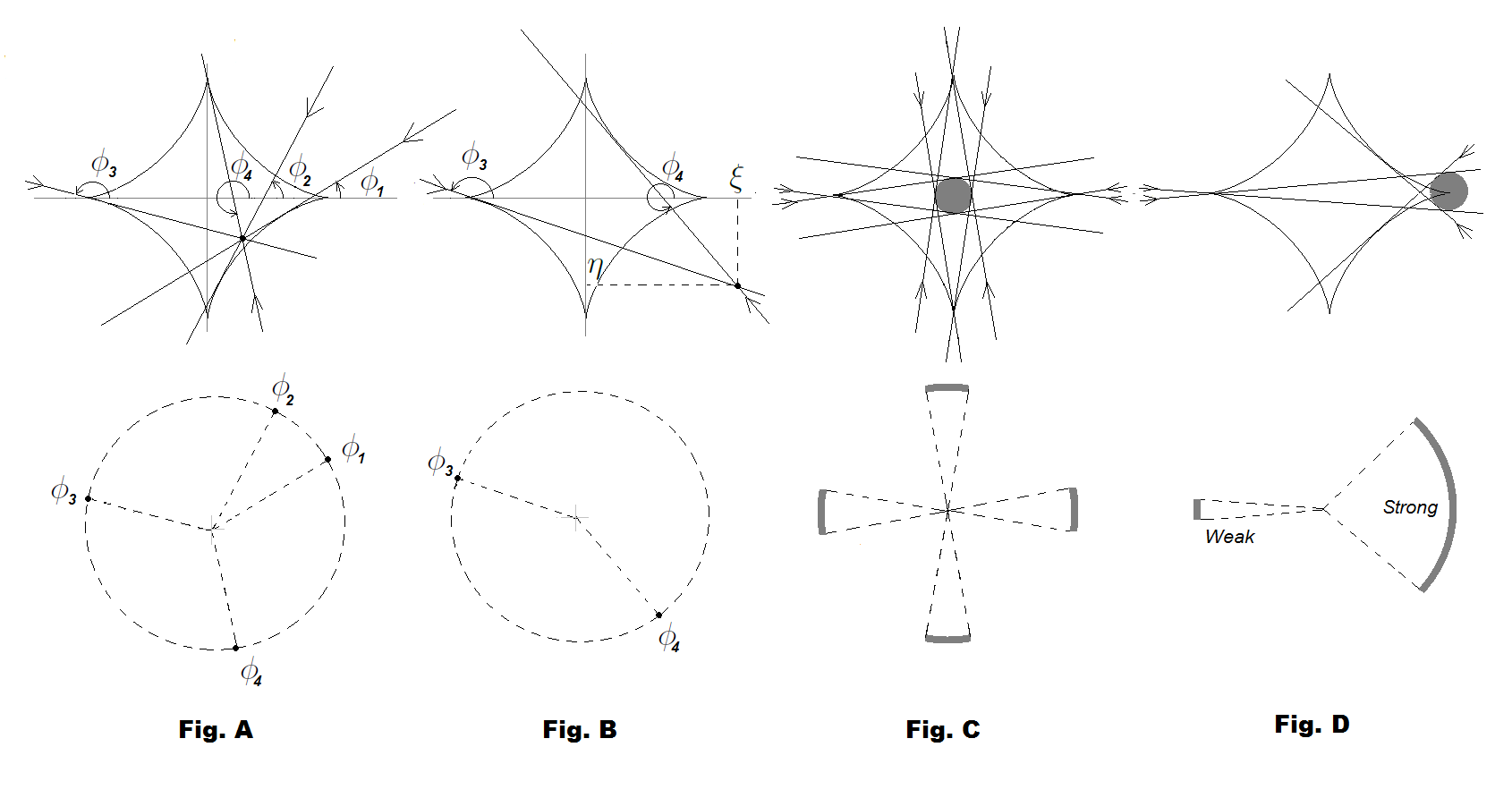}
\caption{ Top: Observer plane. Bottom: Lens plane. Arrows on the tangents indicate directions of the coming rays.
{\bf Fig.A}: Four point images of a point source viewed by an observer from an interior of the astroid.
{\bf Fig.B}: Observer is outside the astroid (two point images).
{\bf Fig.C}: Spans of the (``limb") images of a small (centered at the $z$-axis) disc source viewed by an observer at the $z$-axis (``Einstein Cross").
{\bf Fig.D}: Spans of the strong and weak images of the same small disc source viewed by an observer near the cusp: Maximal angular span of the ``strong limb" is approximately proportional to the cubic root of the ratio between the size of the heliocentric projection of the source to the observer plane and the size of the astroid (while the span of the ``weak limb", as well as limbs of Fig.C, is linearly proportional to the above ratio). Note that image of the disc source does not form a full ring if the apparent size of the disc is smaller than $d_{\rm astroid}/2Z$.}
\label{Images}
\end{figure}

When the deviation of the observer from the $z$-axis is much smaller than $b$, i.e. when $|\xi|\ll 1$ and $|\eta|\ll 1$, we have
$$
\rho=1+\delta, \quad \delta\ll 1 .
$$
Then, from (\ref{xy}) it follows that $\xi+i\eta=2\delta e^{i\phi}-2\epsilon e^{3i\phi}$. Eliminating $\delta$ from the last expression we get
\begin{equation}
\xi\sin\phi-\eta\cos\phi=2\epsilon\sin2\phi .
\label{xietaphi}
\end{equation}
The solutions $\phi$ of (\ref{xietaphi}) are the ``Einstein ring" coordinates of images of the point $(\xi, \eta)$. It is not difficult to see that those are angles between the $\xi$-axis and the tangents to the astroid (\ref{caustic}) drawn from the point $(\xi, \eta)$ (see Figure \ref{Images}). When the point $(\xi, \eta)$ is inside the astroid, equation (\ref{xietaphi}) has four solutions. Otherwise it has two solutions. The magnification of the $j$th image $\mu_j$ is inversely proportional to the distance from the observer to the corresponding tangency point on astroid (see also Appendix 1).

In the $\epsilon=0$ case the lens equation has two solutions when $\xi^2+\eta^2\not=0$ or infinite number of solutions forming a unit circle when $\xi=\eta=0$. Since the number of solutions does not exceed four in the non-degenerate case, the image of a source of small (wrt astroid) apparent size never forms the ring \footnote{Obviously, this does not imply that images of bigger sources, such as e.g. exo-planets of interest, do not form rings.} if $\epsilon\not=0$ (see Figure \ref{Images}).

Returning to the $\epsilon=0$ case we note that the caustic (\ref{caustic}) degenerates to the focal line $\xi=0,\eta=0$ and the deflected beams converge towards the $z$-axis at the ``Einstein" angles \footnote{For a beam grazing the edge of the sun $\alpha_E\approx 0.85 \cdot 10^{-5} $ rad $\approx 1.75$  arcsec.} $|\alpha|=\alpha_E(Z)$
\begin{equation}
\alpha_E(Z)=\sqrt{\frac{2r_g}{Z}} .
\label{parameters2}
\end{equation}
The ``on-axis" observer ``sees" the whole critical line. In other words, the rays are coming from the circle of radius $b$ in the lens plane towards an observer at $X=Y=0$ in the $z=Z$ plane. Therefore, $b > R_0$, where $R_0$ is the sun radius. The distance from the sun to the closest focal point is determined by the condition $b=R_0$ and equals $Z_{\rm min}=R_0^2/(2r_g) \approx 550$ AU.

In the wave optics, the maximal spatial resolution of the spherical lens in the neighborhood of the focal line is restricted by the radius of diffraction (radius of the Airy disc), which is of order $\lambda/\alpha_E$, where $\lambda$ is the light wavelength \cite{HS1},\cite{HS2},\cite{T}, \cite{TT}:  The circularly symmetric diffraction pattern of a point source oscillates in the radial direction and its intensity reaches maximum at the $z$-axis. The spatial scale of the oscillations is of order of the diffraction radius. For $\lambda =1\mu {\mathrm m}$, at the position of the closest observation $Z \approx 550$AU, this radius is about decimeters.

On the other hand,  from eqs.(\ref{parameters1}, \ref{r}, \ref{epsilon}, \ref{caustic}) it follows that the non-spherical model produces an astroid caustic of the diameter \footnote{Also note that $\epsilon=\epsilon_{\rm max} \sin^2(\beta) Z_{\rm min}/Z$, where for the sun $\epsilon_{\rm max}\approx 10^{-7}$}
\begin{equation}
d_{\rm astroid}=8\epsilon b=d_{\rm max}\sqrt{\frac{Z_{\rm min}}{Z}}\sin^2\beta, \quad d_{\rm max}=4I_2 R_0,
\label{diameter}
\end{equation}
which reaches up to $d_{\rm max}\approx 5.6\times 10^2$ meters when $\beta\to\pi/2$ and $Z\to Z_{\rm min}$. Thus, effects of oblateness clearly lead to significant changes of the diffraction pattern of the point source when it moves from the sun polar axis to the equatorial plane. Indeed, the maximum of magnification is now reached in a neighborhood of the astroid, where the geometrical optics magnification diverges, thousands of diffraction radiuses away from the $z$-axis, which is now nonsingular.

We note that, as follows from (\ref{diameter}),
the size of the astroid is proportional to $Z^{-1/2}$, i.e. the size varies slowly with the distance from the sun. For example, the maximal astroid diameter is about $400$ meters at 1000AU, while sizes of the heliocentric projection of possible objects of observation are about several kilometers across.

Before going to the detailed wave optics computations in the next section, it is worthwhile to mention some heuristic arguments explaining significant difference in the maximal EM energy flux amplification in the $\beta=0$ and $\beta=\pi/2$ cases for point source and small wavelengths: In the $\epsilon=0$ case (spherical lens) the lens plane image produced by a small distant source consists either of one single ring or
two opposite arc-shaped ``limbs" (one limb inside and another outside  the critical curve $\rho=\rho_c=1$). Whether the image is a ring or limbs, as well as the size of limbs, depends on the source size and the observer position $X,Y$ in the $z=Z$ plane (the further away the observer is from the $z$-axis, the smaller are the limbs).

In difference from the symmetric case, in the case when the lens caustic is the astroid (\ref{caustic}) the image of a small source \footnote{By ``small" we mean the source whose heliocentric projection to the observer plane has dimensions which are much more smaller than the size of the astroid.} never forms the whole ring (see Figure 2). Such an image consists of two to four disjoint small limbs, some of them being weak and some strong, depending on the observer position wrt caustic.
When the source size goes to zero, the limbs become a set of (two to four) points. Considering the principal Fresnel zones
\footnote{Since S (modulo a $Z$-dependent term) equals the optical path, point $(x, y)$ belongs to the principal zone if $|S(X,Y; x,y)-S(X,Y; x_i,y_i)|<\lambda/2$, where $(x_i,y_i)$ are coordinates of an image.}
around these points one can explain decrease of maximal magnification in the wave optics.

\begin{figure}[htb]
  \centering
  \includegraphics[width=105mm]{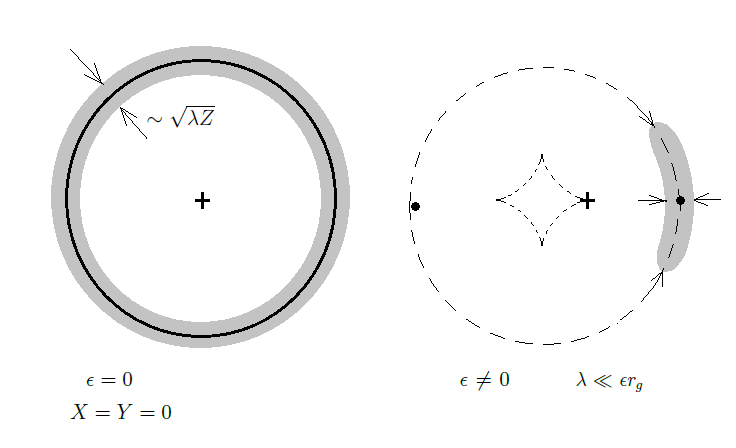}
  \caption{Schematic illustration of contraction of principal Fresnel zones (gray color) towards geometrical optics images of the point source as $\lambda \to 0$. The images are shown by the black circle for $\epsilon=0$, $X=Y=0$ (left) or by fat black dots for $\epsilon \not =0 $ (right). Directions of contraction are indicated by arrows. An observer is denoted by the cross. The critical line and caustic are schematically drawn by dashed lines. In the symmetric case (left) the maximal zone corresponds to the on-axis position of an observer. The zone is an annulus that contracts towards the circle as $\lambda\to 0$. In the $\epsilon\not=0$, $\lambda \ll \epsilon r_g$ case (right) the maximal zones correspond to positions near cusps. In this case a zones contract both radially and tangentially towards point images as $\lambda\to 0$.
  The angular span of the maximal zone in the $\epsilon\not=0$, $\lambda\ll\epsilon r_g$ case is proportional to $(\lambda/\epsilon r_g)^{1/4}$, since e.g. near the cusp corresponding to $\phi=0$: $\phi=\delta\phi, \rho=\rho_c(0)+\delta\rho$ and the variation of the Fermat potential for an observer at the cusp equals $\delta S \approx 2r_g\delta \rho^2 + \epsilon r_g\left(3\delta\rho\delta\phi^2-3\delta\rho^2\delta\phi^2+\delta \phi^4/2\right)$.  The radial dimension of zones is proportional to $(\lambda Z)^{1/2}$.}  \label{Fresnel}
\end{figure}

In more details (see Figure \ref{Fresnel}): the above zones have form of ``limbs", whose dimensions depend on the size of caustic $\sim \epsilon b$, on the diffraction radius $\sim \lambda/\alpha_E$ and on a position of the observer: It is easy to see that the thickness (i.e. radial dimension) of the limb is approximately the same in symmetric and non-symmetric cases when $\epsilon$ is small.

However, in the spherically symmetric case even a point source can produce a circular geometrical optics image, and the maximal zone spans the whole circle even if $\lambda\to 0$ (thickness of maximal zone goes to zero while its angular span always equals $2\pi$). In contrast to the symmetric case, in the $\epsilon\not=0$ case the geometrical optics image of a point is always a point and therefore the maximal principal Fresnel zone contracts to a point as $\lambda\to 0$. This leads to decrease of the maximal EM energy flux amplification in comparison with the symmetric case
\footnote{ An estimate of the maximal magnification, based on evaluation of dimensions of the Fresnel zone corresponding to the cusp of the astroid (\ref{caustic}) was presented in \cite{Eshleman} using an analogy with focusing by atmosphere of an oblate planet \cite{ETF}. There, the angular span (``horizontal dimension" in terminology of \cite{ETF}) of the maximal Fresnel zone was estimated by  using the following assumption (Section 6 of \cite{ETF}): The variation, along the half-span of the zone, of distance between the cusp and the involute of the astroid equals the ``vertical" dimension (i.e. thickness) of the zone. From this it follows that the angular span of the zone is proportional to $(\lambda/\epsilon^2 r_g)^{1/8}$ as $\lambda\to 0$, while from direct evaluation of the variation of the Fermat potential $S$ it follows that the span $\sim(\lambda/\epsilon r_g)^{1/4}$.}
(since the thicknesses of zones have practically the same dependence on $\lambda$ when $\epsilon$ is small).

\section{ Diffraction Optics and Gravitational PSF}

The geometrical  theory of diffraction provides algorithms for finding near-caustic intensity from its geometric optics asymptotics \cite{LL}, \cite{BK}. For caustics of the type (\ref{caustic}) the near-field is expressed through the Airy function for caustic folds (see e.g. \cite{LL}, \cite{SEF}, \cite{ND}), and through the Pearcey integral near caustic cusps (see e.g. \cite{BK}, \cite{P}). In cases of focal lines the near-field is expressed through Bessel functions \cite{BK}, \cite{ND}.

Application of these algorithms to our problem is presented in the Appendix 1.

Below, we derive the gravitational point spread function as a single one-dimensional integral having all the above mentioned limits.

The change of polarisation angles of light in weak gravitomagnetic fields is of a post-post Newtonian order and can be neglected (see eg \cite{S}, \cite{KM} and references therein). The deflection angles are also small and space is asymptotically flat, so we can apply the scalar Huygens-Fresnel principle (Fresnel-Kirchhoff diffraction formula): In the thin lens, short wavelength approximation ($Z\gg R_0 \gg r_g$, $\lambda\ll r_g$, $Z_{\rm min}/r_g\approx 3\times10^{10}$) the diffraction field in vicinity of the $z$ axis is a sum of contributions by spherical waves propagating from the lens plane $z=0$ with the phase delays corresponding to the sum of the gravitational and geometric delays. This total time delay equals $S/c$, where $S$ is given by (\ref{lens2}). Then, the complex amplitude of the electromagnetic (EM) field at the observer position equals (up to the phase factor $e^{ikZ}$)
\begin{equation}
u=\frac{k}{2\pi Z}\int e^{ikS}dxdy, \quad k=2\pi/\lambda .
\label{HuygensFreshnel}
\end{equation}
The intensity magnification, i.e. amplification of the EM energy flux at the point $X,Y$ of the observer plane, equals the square of the absolute value of the amplitude \footnote{Dropping common phase factors, e.g. such as $e^{ikZ}$ in (\ref{HuygensFreshnel}) etc, does not affect the value of the above gravitational magnification.  Therefore, for simplicity, we will perform our computations of $u$ modulo common phase factors. Note that factor $e^{ikZ}$ cannot be dropped in (\ref{HuygensFreshnel}) if one considers a combination of the gravitational lens with an optical device (see Section 5).}
$$
\mu = |u|^2.
$$
This function of $X,Y$ ($\mu$ also depends on $Z$, $\epsilon$ and $k$) is a point spread function (PSF) of a gravitational lens only, so we call it gravitational PSF or GPSF (the PSF of a combination of the gravitational lens and a telescope is discussed in Section 5).

For a detailed derivation and justification of the Fresnel-Kirchhoff diffraction integral (\ref{HuygensFreshnel}) see e.g. \cite{SEF} or \cite{ND}. In (\ref{HuygensFreshnel}), the geometrical optics magnification (\ref{mu}) is recovered as $\lim_{k\to\infty}|u|^2$. When $kr_g\gg 1$ and the point on the observer $X,Y$-plane is far away from caustic, the integral (\ref{HuygensFreshnel}) can be expressed in a simple manner through the geometrical-optics data as (see e.g. \cite{ND}, \cite{SEF})
\begin{equation}
u(X,Y) = \sum_j \sqrt{|\mu_j|} e^{i(kS_j-\pi n_j/2)}.
\label{eS}
\end{equation}
Here the sum is taken over the number of images in the lens plane, $\mu_j$ is the geometrical optics magnification of the $j$th image, $S_j=S(X,Y; x_j,y_j)$ is the extremal value of the Fermat potential (\ref{lens2}) for the $j$th image and $n_j=0,1,2$ corresponds to $x_j,y_j$ being the minimum, saddle and the maximum point respectively. The above approximation breaks down in the neighborhood of caustic, the case we are mainly interested in. So, one has to either evaluate (\ref{HuygensFreshnel}) exactly or to apply suitable asymptotic methods.

Up to a common ($X,Y,Z$-dependent) phase factor
\begin{equation}
u(\xi,\eta)=\frac{q}{2\pi}\int_{R_0/b}^\infty d\rho\int_0^{2\pi} \rho e^{ iqV} d\phi, \quad V= \frac{\rho^2}{2}-\rho(\xi\cos\phi+\eta\sin\phi)-\psi(\rho,\phi) ,
\label{A}
\end{equation}
where $q$ is the dimensionless wavenumber
\begin{equation}
q=2 k r_g=\frac{4\pi r_g}{\lambda} .
\label{q}
\end{equation}
For $\lambda=10^{-6}$m, $q\approx 3.7\times 10^{10}$.

The main purpose of this section is the direct numerical evaluation of the 2d integral (\ref{A}). Before presenting the numerical results we would like to make several remarks: \\

{\bf Remark 1}: It is worthy to note that in the $\epsilon=0$ case, the exact 2d integration in (\ref{A}) is possible:
\begin{equation}
u=\frac{q}{2\pi}\int \rho d\rho \int_0^{2\pi} e^{iq\left[\frac{\rho^2}{2}-\log(\rho)-\rho(\xi\cos\phi+\eta\sin\phi)\right]}d\phi = q\int \rho d\rho e^{ iq\left[\frac{\rho^2}{2}-\log(\rho)\right]}J_0\left(q\rho\sqrt{\xi^2+\eta^2}\right),
\label{Exact}
\end{equation}
where $J_0$ is the zero-order Bessel function. After integration in $\rho$ one can express $\mu$ in terms of the confluent hypergeometric function (see e.g. \cite{ND})
$$
\mu=|u|^2=\frac{\pi q}{1-e^{-\pi q}}\left|{_1F_1}\left(iq/2,1;iq(\xi^2+\eta^2)/2\right)\right|^2 .
$$
In the short-wavelength limit $q\gg 1$ and when the argument $iq(\xi^2+\eta^2)/2$ of ${_1F_1}$ is small, i.e.
\begin{equation}
\sqrt{\xi^2+\eta^2}\ll \frac{1}{\sqrt{q}}, 
\label{small_lim}
\end{equation}
the hypergeometric function ${_1F_1}$ degenerates to the zero-order Bessel function (see e.g. \cite{ND}, \cite{TT})
$$
\mu = \pi q J_0^2\left(q\sqrt{\xi^2+\eta^2}\right) .
$$
The maximum $\mu=\mu_0$ of the GPSF is reached at the focal line $\xi=\eta=0$ and equals
\begin{equation}
\mu_0=\pi q =\frac{4\pi^2 r_g}{\lambda} .
\label{mu_0}
\end{equation}

{\bf Remark 2}: Condition (\ref{small_lim}) is, in fact, a condition of validity of the stationary phase integration at $\rho=1$ in the last integral in (\ref{Exact}). Indeed, the stationary phase approximation can be applied in (\ref{Exact}) when the width of the stationary phase region $\delta \rho \sim 1/\sqrt{q}$ is much more smaller than the scale of oscillations of the Bessel function $\delta \rho \sim 1/(q\sqrt{\xi^2+\eta^2})$, which leads to (\ref{small_lim}).

{\bf Remark 3}: Condition (\ref{small_lim}) will be encountered in the next section, when a similar type of the stationary phase integration will be performed for the general case $\epsilon\not=0$: As follows from (\ref{A}), in the general situation
\begin{equation}
u=q \int \rho e^{ iq\left[\frac{\rho^2}{2}-\log(\rho)\right]} F(\epsilon q/\rho^2, q\xi\rho, q\eta\rho) d\rho ,
\label{integrate_phi}
\end{equation}
where the function of three variables $F(\cdot, \cdot, \cdot)$ is defined as follows  \footnote{$F(\chi, \kappa, \nu) $ degenerates to $J_0$ in two special cases: 1) $\chi=0$ (see Remark 1) and 2) $\kappa=\nu=0$ (see eq. (\ref{on_axis})). }
$$
F(\chi, \kappa, \nu)=\frac{1}{2\pi}\int_0^{2\pi}e^{\chi\cos2\phi-\kappa\cos\phi-\nu\sin\phi}d\phi .
$$
Similarly to the $\epsilon=0$ case, the stationary phase integration can be performed  at $\rho=1$  when all the arguments of $F$  in (\ref{integrate_phi}) are much smaller than $\sqrt{q}$, i.e. when
\begin{equation}
\epsilon \ll \frac{1}{\sqrt{q}}
\label{small_epsilon}
\end{equation}
and (\ref{small_lim}) holds. This will be demonstrated in detail in the next section. \\

Consider now the general case, i.e. the one when $(\ref{small_lim})$ is not necessarily true. Below we will proceed with the main subject of the present section, estimating the 2d integral (\ref{A}) numerically without any assumptions.

Since $q\gg1$ (e.g. $q\approx 3.7\times 10^{10}$ for $\lambda=10^{-6}$m) one can reduce (\ref{A}) to a one-dimensional integral in $\phi$ by the stationary-phase integration \footnote{For $q\gg 1$, $\int e^{iqf(x)} xdx = x_s\sqrt{\frac{2\pi i}{qf''(x_s)}}e^{iqf(x_s)}\left(1+{\cal O}\left(\frac{1}{q}\right)\right)$, where $ f'(x_s)=0$} in $\rho$ at fixed $\phi$: The stationary phase integration in $\rho$ produces a relative ${\cal O}(1/q)$ error, which is negligible for wavelengths of interest.

First, one should find the ``stationary phase line" $\rho = \rho_{\rm st}(\xi, \eta; \phi)$ a such that \footnote{It is not difficult to see that solutions of the lens equation (\ref{xy}) lie on this line. When the observer is far away from the caustic, the stationary phase integration in $\phi$ can be also performed around these points together with the above integration in $\rho$. Such a double stationary phase integration results in (\ref{eS}). This is not the case when the observer is in a neighborhood of the caustic since the second tangential derivative of $V$ vanishes on the critical line and one has to apply other methods for computing the integral in $\phi$.}
\begin{equation}
\left(\frac{\partial V}{\partial \rho}\right)_{\rho=\rho_{\rm st}} =
\rho_{\rm st}-\frac{1}{\rho_{\rm st}}-\xi\cos\phi-\eta\sin\phi-\frac{2\epsilon\cos2\phi}{\rho_{\rm st}^3}=0 .
\label{Vrho}
\end{equation}
Then, up to a common phase factor
\begin{equation}
u=\sqrt{\frac{q}{2\pi}}\int_0^{2\pi} \left(\frac{\partial^2 V}{\partial \rho^2}\right)^{-1/2}_{\rho=\rho_{\rm st}(\phi)}\rho_{\rm st}(\phi) \exp\left[iqV(\rho_{\rm st}(\phi),\phi)\right] d\phi.
\label{ustat}
\end{equation}
From (\ref{Vrho}) we obtain \footnote{Eq.(\ref{Vrho}) has two solutions: $\rho_1(\phi)=\rho_{\rm st}(\phi)$ and $\rho_2(\phi)=-\rho_{\rm st}(\phi+\pi)$. A positive solution has to be chosen, since the integration in $\rho$ in (\ref{A}) is performed for $\rho >0$. Note, however, that permutation of the solutions only reverses the sign of integral (\ref{ustat}) since the both solutions parametrize differently the same curve.} that up to ${\cal O}(\epsilon^2)$
\begin{equation}
\rho_{\rm st}(\xi,\eta;\phi)=\frac{\sqrt{4+\tau^2}+\tau}{2}+\frac{8\epsilon\cos2\phi}{(\sqrt{4+\tau^2}+\tau)^2\sqrt{4+\tau^2}}, \quad \tau=\tau(\xi,\eta;\phi)=\xi\cos\phi+\eta\sin\phi
\label{rhoepsilon}
\end{equation}
Therefore the GPSF (intensity magnification) expresses through the one-dimensional integral in $\phi$:
\begin{equation}
\mu=|u|^2=\pi q\left|F\right|^2=\mu_0\left|F\right|^2 ,
\label{u}
\end{equation}
\begin{equation}
F=\frac{1}{\sqrt{2}\pi}\int_0^{2\pi} \left(1+\frac{1}{\rho_{\rm st}^2}+\frac{6\epsilon\cos2\phi}{\rho_{\rm st}^4}\right)^{-1/2}\rho_{\rm st} \exp iq\left[\frac{\rho_{\rm st}^2}{2}-\log\rho_{\rm st}-\tau\rho_{\rm st}+\frac{\epsilon\cos2\phi}{\rho_{\rm st}^2}\right] d\phi ,
\label{Inteps}
\end{equation}
where two functions of $\phi$: $\tau=\tau(\xi, \eta; \phi)$ and $\rho_{\rm st}=\rho_{\rm st}(\xi, \eta; \phi)$ are given in (\ref{rhoepsilon}).

\begin{figure}[htb]
  \centering
  \includegraphics[width=175mm]{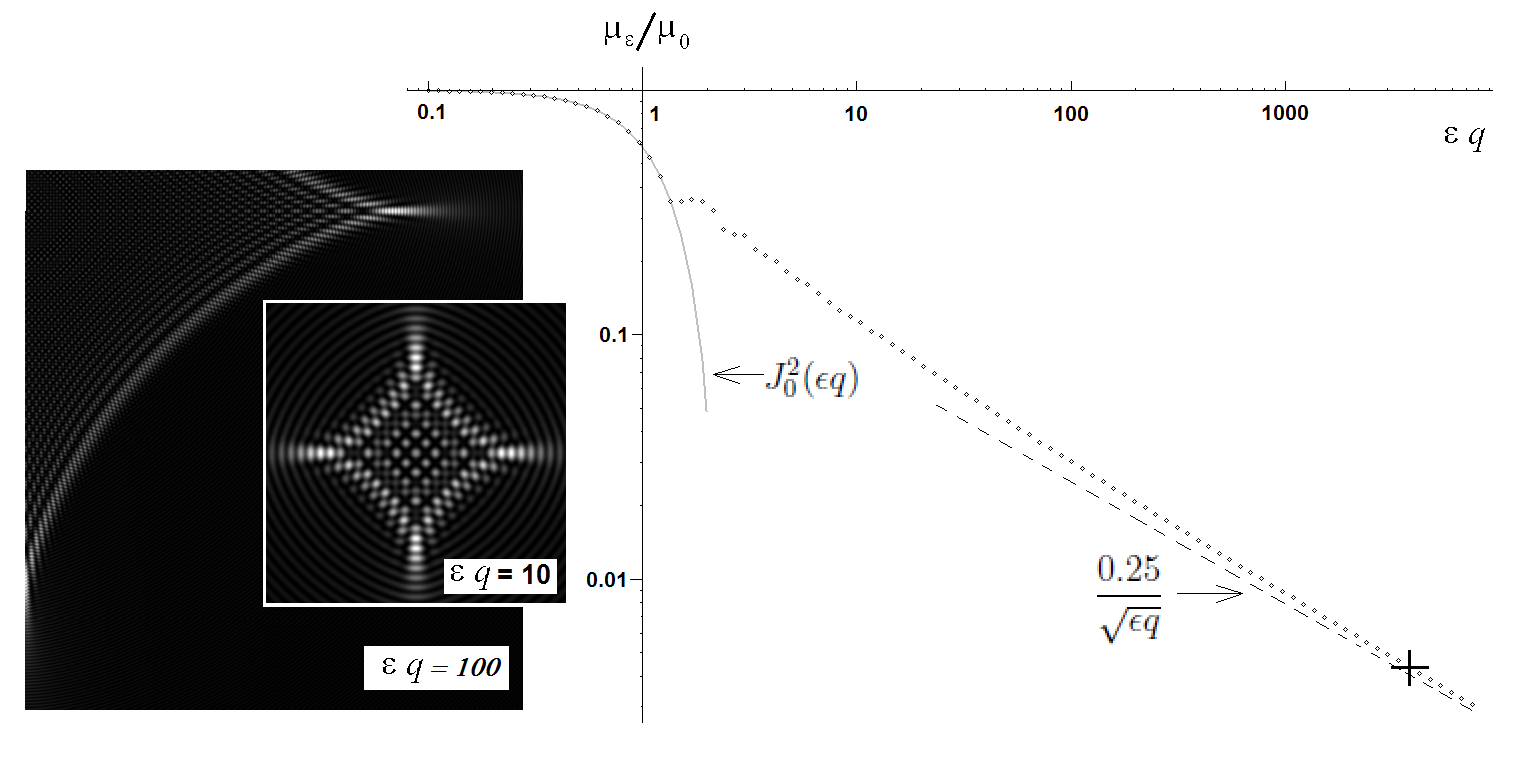}
  \caption{
  Left Figure: Normalized diffraction patterns for $q=3.7\times10^{10}$, $\epsilon q = 10$ and $\epsilon q = 100$. The former image ($q\epsilon=10$) is $3\times$ zoomed in wrt to the latter. Right Figure: The $\log-\log$ plot of $\mu_\epsilon/\mu_0$ as a function of $\epsilon q$. Numerical evaluation is shown by the point plot. High $\epsilon q$ asymptotics (\ref{Num}) is shown by the dashed line. Low $\epsilon q$ approximation (see eq.(\ref{on_axis})) is shown by gray solid line. The point corresponding to observations in the equatorial plane of the sun at $Z=550$ AU and $\lambda=10^{-6}$m is marked with a cross. At this point $\epsilon q \approx 3.7\times 10^3$ and $\mu_\epsilon/\mu_0 \approx 4\times 10^{-3}$. Computations are performed for $q=3.7\times 10^{10}$ that corresponds to $\lambda\approx 10^{-6}$m. Small parameter $\epsilon$ varies from $\epsilon\approx 2.7\times 10^{-12}$ to $\approx 2.0\times 10^{-7}$.  Error of numerical integration $\delta\mu_\epsilon/\mu_\epsilon$ does not exceed $\approx 1$ percent.
  } \label{muqe}
\end{figure}

The above integral can be particularly easy taken when $\epsilon=0$, $\xi=\eta=0$, giving the value of $\mu_0$, obtained earlier (\ref{mu_0}). In the general case, the integral (\ref{Inteps}) should be taken numerically.

Now, we present results of the direct numerical computation of the GPSF (\ref{u},\ref{Inteps}): Figure \ref{muqe} shows a ratio of the maximal intensity $\mu_\epsilon$ at $\epsilon\not=0$
$$
\mu_\epsilon:=\max_{X,Y}\mu(X,Y; q, \epsilon)
$$
to that of the symmetric case $\mu_0=\pi q$ for different values of the parameter $q\epsilon$. The computations presented at Figure \ref{muqe} are performed at fixed $q\approx 3.7\times 10^{10}$ (which corresponds to $\lambda\approx 1\mu$m) for different values of $\epsilon$. However, the numerical results (as well as the analysis in the next section) show that $\mu_\epsilon/\mu_0$ approximately depends only on $\epsilon q$ when (\ref{small_epsilon}) holds.

As seen from Figure \ref{muqe} one naturally recovers the unit ratio for $q\epsilon\to 0$
$$
\frac{\mu_\epsilon}{\mu_0} \to 1, \quad q\epsilon\to 0 .
$$
(More precisely $\mu_\epsilon/\mu_0 \approx J_0^2(\epsilon q)$ when $\epsilon q<\approx 1.4$. See eq. (\ref{on_axis}) and the end of this section).

On the other hand
\begin{equation}
\frac{\mu_\epsilon}{\mu_0}\approx \frac{0.25}{\sqrt{\epsilon q}}
\label{Num}
\end{equation}
when $q\epsilon \gg 1$. In this case four equal maxima of the intensity magnification are symmetrically situated on the $X$, $Y$ axes: two at the $X$-axis and another two at $Y$-axis in the caustic interior close to four cusps $(X,Y)=(\pm4 b\epsilon, 0)$, $(X, Y)=(0, \pm 4b\epsilon)$. When $q\epsilon$ is large, the distance between the cusp and the neighboring maximum is proportional to $\epsilon b/\sqrt{\epsilon q}$, while value of $\mu$ at the cusp is proportional to $\mu_\epsilon$ (i.e. $\mu_{\rm cusp}\approx 0.5 \mu_\epsilon$).

The GPSF oscillates and the amplitude of oscillations grows as one moves towards the neighborhood of the caustic folds/cusps. The amplitude falls off similarly to the $\epsilon=0$ case and the pattern becomes more and more radially symmetric, as one moves far away from caustic in the exterior direction (which is in agreement with (\ref{eS})).

As $q\epsilon$ becomes smaller the four global maxima approach (discontinuously, see below) the center $X=Y=0$. The diffraction pattern becomes circularly symmetric at $q\epsilon=0$.

It is useful to introduce the parameter $\chi=\epsilon q$:
\begin{equation}
\chi=\epsilon q=\frac{\lambda_0}{\lambda}\frac{Z_{\rm min}}{Z}\sin^2\beta, \quad \lambda_0=2\pi I_2 r_g .
\label{chi}
\end{equation}
As follows from (\ref{Num})
$$
\frac{\mu_\epsilon}{\mu_0}\approx \frac{0.25}{\sqrt{\chi}} = \frac{0.25}{|\sin \beta|}\sqrt{\frac{\lambda}{\lambda_0}\frac{Z}{Z_{\rm min}}}, \quad {\rm when} \quad \chi \gg 1 .
$$
In the case of the sun $\lambda_0 \approx 3.7\times 10^{-3}$m, i.e. the effects due to quadrupole moments of the sun can be noticeable already at the far-infrared part of the EM spectrum. For $\lambda = 10^{-6}$m, the parameter $\chi$ can be as large as $\approx 3.7\times 10^3$. For these wavelengths the maximum magnification of the energy flux from the point source can decrease up to several orders of magnitude when the source goes from the polar axis of the sun $\beta=0$ towards its equatorial plane $\beta=\pi/2$.

Concluding this section we would like to mention that (as can be seen from Figure (\ref{muqe})) $\mu_\epsilon/\mu_0$ is a non-smooth function of $\chi$: its derivative wrt $\chi$ jumps at certain points (e.g. the first jump occurs at $\chi \approx 1.4$, the second at $\chi\approx 2.5$ etc). A jump takes place every time when some of the local maxima of $\mu(X,Y)$ become the global ones: For example, the extremum at the center $X=Y=0$ ($\mu\vert_{X=0,Y=0}\approx\mu_0 J_0^2(\chi)$, see eq. (\ref{on_axis})) is the global maximum when $\chi<\approx 1.4$. At $\chi\approx 1.4$ this maximum becomes smaller than four maxima at the distance $\approx 0.54 \lambda/\alpha_E$ from the center. As $\chi$ increases, positions of new global maxima change continuously until the next jump at $\chi\approx 2.5$ etc.  \\

In the next section we give an analytic explanation of the numerical results obtained.

\section{Diffraction Optics: Limiting Cases}

For analytic description of the above numerical result we expand the argument of exponential in (\ref{Inteps}) in $\tau$ and $\epsilon$ (we recall that $\tau=\frac{X}{b}\cos\phi+\frac{Y}{b}\sin\phi$, see eq. (\ref{rhoepsilon}))
\begin{equation}
F=\frac{1}{2\pi}\int_0^{2\pi} G {e^{iqU}}d\phi ,
\label{F2}
\end{equation}
where, to the second order in $\tau$ and $\epsilon$
\begin{equation}
U=\frac{1}{2}-\tau+\epsilon\cos2\phi-U_2, \quad U_2=\left(\frac{
\tau}{2}+\epsilon\cos2\phi\right)^2
\label{U2}
\end{equation}
and to the first order in $\tau$, $\epsilon$
\begin{equation}
G=1+\frac{3}{4}\tau .
\label{G1}
\end{equation}
We are interested in the short wavelength limit $q\gg1$ of (\ref{F2}). The $U_2$-term in (\ref{F2}, \ref{U2}) can be neglected if $q\left|U_2\right|\ll \pi$. This condition holds when
\begin{equation}
|\tau|\ll \frac{2}{\sqrt{q}}, \quad \epsilon \ll \frac{1}{2\sqrt{q}}, \quad q \gg 1 .
\label{small}
\end{equation}
The above conditions have been already encountered in (\ref{small_lim}), (\ref{small_epsilon}).

Since $|\tau|\ll2/\sqrt{q}$ and $q\gg1$, the $\tau$-term in (\ref{G1}) can be dropped. Therefore, provided (\ref{small}) holds, up to the constant phase factor
\begin{equation}
F=\frac{1}{2\pi}\int_0^{2\pi}e^{iq(\epsilon\cos2\phi-\tau)}d\phi=\frac{1}{2\pi}\int_0^{2\pi}e^{iq(\epsilon\cos2\phi-\xi\cos\phi-\eta\sin\phi)}d\phi .
\label{F0}
\end{equation}
Rewritten in terms of $X$ and $Y$, the condition imposed on $\tau$ in (\ref{small}) is $\left|X\cos\phi+Y\sin\phi\right|\ll 2 b/\sqrt{q}$, i.e.
\begin{equation}
R\ll R_{\rm v} = \frac{2b}{\sqrt{q}}=\sqrt{\frac{2}{\pi}}\sqrt{\lambda Z}, \quad R:=\sqrt{X^2+Y^2},
\label{R_validity}
\end{equation}
where $R$ is the distance between the observer and the $z$-axis. For $\lambda \sim 10^{-6}$m (i.e. for $q\sim 10^{10}-10^{11}$), the radius $R_{\rm v}$ is about 10 kilometers, while the maximum possible radius of the caustic is about $300$ meters, so the above condition of validity of (\ref{F0}) clearly holds in the region of the interest \footnote{The ratio of the radius $R_{\rm v}$ to the caustic radius $4\epsilon b$ equals $1/(2\epsilon\sqrt{q})$, which leads to the condition $\epsilon\ll 1/\sqrt{q}$. Since for the sun $\epsilon$ is at most about $10^{-7}$, this condition clearly holds in our case.}.

Therefore, to get the amplitude of EM field one can integrate over the circle $\rho = 1$ in the lens plane provided (\ref{R_validity}) holds. This happens due to the fact that the ``optical path" $S$ is extremal \footnote{The second tangential derivative of $S$ also vanishes on the critical line.} on the ray trajectories (see eq. (\ref{lens1})) and small deformations of the integration contour do not significantly change contribution from the ``monopole part" $S_0$ of $S=S_0+\epsilon S_1$ when (\ref{R_validity}) holds. Since the contour deformations are of order of $\epsilon$, the error in the quadrupole contribution $\epsilon S_1$ is of order of $\epsilon^2$, which is also negligible. In other words, the width of the stationary phase integration region ("thickness of Fresnel zone", see Figure \ref{Fresnel}) significantly exceeds the deviation of the integration contour  $\rho=\rho_{\rm st}(\xi,\eta,\phi)$ from the unit circle when (\ref{R_validity}) is true.

It follows from (\ref{F0}) and (\ref{u}) that $\mu(X,Y; q, \epsilon)/\mu_0(q)$ is essentially a function of three variables
$$
\frac{\mu}{\mu_0}=f\left(\epsilon q, \frac{qX}{b},\frac{qY}{b}\right)
$$
when (\ref{small_lim}) and (\ref{small_epsilon}) hold. Therefore our results are scalable. For instance, $\mu_\epsilon/\mu_0$ evaluated for equatorial observations  at $Z=550$AU for $\lambda=1\mu$m is the same as for $\lambda = 0.5\mu$m at $Z=1100$AU. \\

Apart from the situation when (\ref{F0}, \ref{R_validity}) overlaps with the approximation (\ref{eS}), analytical study of (\ref{F0}) can be performed for the three asymptotic cases:


\begin{quote}
{\bf (1)} The ``\textbf{\textit{degenerate}}" case $\chi=q\epsilon\ll 1$, i.e the case of observations in directions that are close to the sun polar axis $|\beta| \ll \sqrt{\frac{\lambda}{\lambda_0}\frac{Z}{Z_{\rm min}}}$ (see eq.(\ref{chi})).

At $Z\sim 1000$AU and $\lambda \sim 10^{-6}$m this corresponds to $\beta$'s that are smaller than a fraction of a degree. These directions cover less than 0.01\% of the celestial sphere.
\end{quote}


\begin{quote}
{\bf (2)} The ``\textbf{\textit{strongly non-degenerate}}" case $\chi=q\epsilon\gg 1$, or equivalently $\epsilon \gg \lambda/r_g$. In this case the scale of the diffraction pattern is much more smaller than the transverse caustic size, which takes place for $|\beta| \gg \sqrt{\frac{\lambda}{\lambda_0}\frac{Z}{Z_{\rm min}}}$.

At $Z\sim 1000$AU and $\lambda\sim 10^{-6}$m, this corresponds to the directions with $\beta$'s bigger than few degrees.
\end{quote}


\begin{quote}
{\bf (3)} ``\textbf{\textit{On-axis}}" magnification, i.e. value of GPSF at $X=Y=0$ and an arbitrary $\chi=\epsilon q$. \\
\end{quote}


{\bf 1.} We start with the first case, the spherical lens. When $\epsilon=0$, (\ref{F0}) degenerates to the zero-order Bessel integral
$$
F=\frac{1}{2\pi}\int_0^{2\pi}e^{-iq(\xi\cos\phi+\eta\sin\phi)}d\phi=J_0\left(q\sqrt{\xi^2+\eta^2}\right) .
$$
Since $\xi=X/b$, $\eta=Y/b$, $q=2kr_g=4\pi r_g/\lambda$ and $b=\sqrt{2r_gZ}$ (see eq.(\ref{parameters1}), (\ref{q})), from (\ref{u}) we get
\begin{equation}
\mu=\pi qJ_0^2(q\sqrt{\xi^2+\eta^2})=\frac{4\pi^2 r_g}{\lambda}J_0^2\left(\frac{2\pi}{\lambda}\sqrt{\frac{2r_g}{Z}}R\right), \quad R=\sqrt{X^2+Y^2} .
\label{mupole}
\end{equation}
Thus the $\epsilon=0$ limit, obtained alternatively in Section 3 by exact 2d integration of (\ref{A})) is recovered. This gives the well-known result \cite{HS1},\cite{HS2}, \cite{ND}, \cite{T}, \cite{TT} for the GPSF of the spherical lens.

The GPSF is circularly symmetric and reaches maximum $\mu_0$ at $R=0$. The radius of the Airy disc (i.e. ``diffraction radius") and the spatial period of the radial Airy pattern is of order of $\lambda/\alpha_E(Z)$.\\

{\bf 2.} Let us now pass to the strongly non-degenerate case $q\epsilon\gg 1$. In this case the size of caustic $\sim \epsilon b$ greatly exceeds the diffraction radius $\sim \lambda/\alpha_E$. First we consider the asymptotic of the GPSF in the cusp neighborhoods, where it reaches the maxima. For convenience we choose the cusp at $X=4\epsilon b,Y=0$, (i.e. at $\xi=4\epsilon, \eta=0$).

We now introduce the cusp-related coordinates $\tilde{\xi}$, $\tilde{\eta}$, such that
$$
\xi=\epsilon\left(4+\frac{2\tilde{\xi}}{\sqrt{2q\epsilon}}\right), \quad \eta=\frac{\sqrt{2}\epsilon\tilde{\eta}}{(2q\epsilon)^{3/4}},
$$
and re-scale the integration angle $\phi\to\varphi$
$$
\phi=\frac{2^{1/4}\varphi}{(q\epsilon)^{1/4}} .
$$
In these coordinates integral (\ref{F0}) rewrites as
\begin{equation}
F=\frac{1}{2\pi(q\epsilon/2)^{1/4}}\int_{-\pi (q\epsilon/2)^{1/4}}^{\pi (q\epsilon/2)^{1/4}}e^{i\tilde{V}(\varphi)} d\varphi ,
\label{FP}
\end{equation}
where, modulo $\varphi$-independent terms, $\tilde{V}$ has the following form
\begin{equation}
\tilde{V}=-\tilde{\eta}\varphi+\tilde{\xi}\varphi^2+\varphi^4+\tilde{V}_1, \quad 
\tilde{V}_1=\frac{\sqrt{2}}{\sqrt{q\epsilon}}\left[\tilde{\xi}\varphi^4 P+\tilde{\eta}\varphi^3 Q+\varphi^6 H\right]
\label{VP}
\end{equation}
Here, $P,Q,H$ are bounded functions ($\max(|P|,|Q|,|H|)\le1/6$) of the single variable $\phi=\frac{\varphi}{(q\epsilon/2)^{1/4}}$, $\phi=-\pi..\pi$.
$$
P=\frac{1-\frac{\phi^2}{2}-\cos\phi}{\phi^4}, \quad Q=\frac{\phi-\sin\phi}{\phi^3}, \quad H=2\frac{\cos2\phi-4\cos\phi+3-\frac{\phi^4}{2}}{\phi^6}, \quad \phi=-\pi..\pi.
$$
Since $q\epsilon\gg 1$, the $\tilde{V_1}$ term in (\ref{VP}) can be neglected when
\begin{equation}
|\tilde{\xi}|\ll \sqrt{q\epsilon}, \quad |\tilde{\eta}| \ll \sqrt{q\epsilon} .
\label{cusplimits}
\end{equation}
Then, provided the above conditions hold, from (\ref{FP}, \ref{VP}) it follows that
$$
F=\frac{1}{2\pi(q\epsilon/2)^{1/4}}{\mathrm {Pe}}(\tilde{\xi},\tilde{\nu}) ,
$$
where ${\mathrm {Pe}}(x,y)$ is the Pearcey integral
\begin{equation}
{\mathrm {Pe}}(x,y)=\int_{-\infty}^\infty e^{i\left(-y\varphi+x\varphi^2+\varphi^4\right)}d\varphi .
\label{Pe}
\end{equation}
It follows from (\ref{u}) that, in terms of unscaled deviation $\tilde{X}=X-4\epsilon b$, $\tilde{Y}=Y$ from the cusp, the near-cusp GPSF equals
\begin{equation}
\mu=\frac{1}{4\pi}\sqrt{\frac{2q}{\epsilon}}\left|{\mathrm {Pe}}\left(\frac{\tilde{X}}{2\epsilon b}(2\epsilon q)^{1/2},\frac{\tilde{Y}}{\epsilon b}(2\epsilon^3 q^3)^{1/4}\right)\right|^2 .
\label{Pearcey}
\end{equation}
The domain of validity  (\ref{cusplimits}) of the above asymptotics rewrites in terms of the unscaled deviations from the cusp as follows
\begin{equation}
\left|\tilde{X}\right|=\left|X-4\epsilon b\right| \ll \epsilon b, \quad \left|\tilde{Y}\right|=\left|Y\right| \ll \frac{\epsilon b}{(q\epsilon)^{1/4}} .
\label{small_cusp}
\end{equation}
We recall that the above conditions are in agreement with (\ref{R_validity}), since, as has been mentioned before, the ``radius of validity" of (\ref{F0}) greatly exceeds maximal possible size of caustic when $\epsilon\ll 1/\sqrt{q}$.

The absolute value of the Pearcey integral $\left|{\mathrm {Pe}}(\tilde{\xi},\tilde{\eta})\right|$ reaches maximum at $\tilde{\xi}\approx -2.02$, $\tilde{\eta} =0$ which is inside the domain of validity (\ref{cusplimits}) of eq. (\ref{Pearcey}). The maximum of GPSF in the $\epsilon q\gg 1$ limit then equals
$$
\mu_\epsilon=\frac{1}{4\pi} \sqrt{ \frac{2q}{\epsilon} }\max \left|{\mathrm {Pe}}\right|^2=\sqrt{\frac{r_g}{2\pi\lambda\epsilon}}\max \left|{\mathrm {Pe}}\right|^2, \quad \max\left|{\mathrm {Pe}}\right|^2\approx 7.02  .
$$
Therefore the ratio of the maximum $\mu_\epsilon$ to that of the spherically symmetric case $\mu_0$ equals
\begin{equation}
\frac{\mu_\epsilon}{\mu_0} = \frac{\sqrt{2}\max \left|{\mathrm {Pe}}\right|^2}{4\pi^2} \frac{1}{\sqrt{q\epsilon}} \approx \frac{0.25}{\sqrt{q\epsilon}}, \quad q\epsilon\gg 1 .
\label{comparison}
\end{equation}
The distance between point of maximum and the cusp equals $\approx 2.02\epsilon b/\sqrt{\epsilon q/2}$. This confirms the numerical results of the Section 3.

\begin{figure}
  \centering
  \includegraphics[width=105mm]{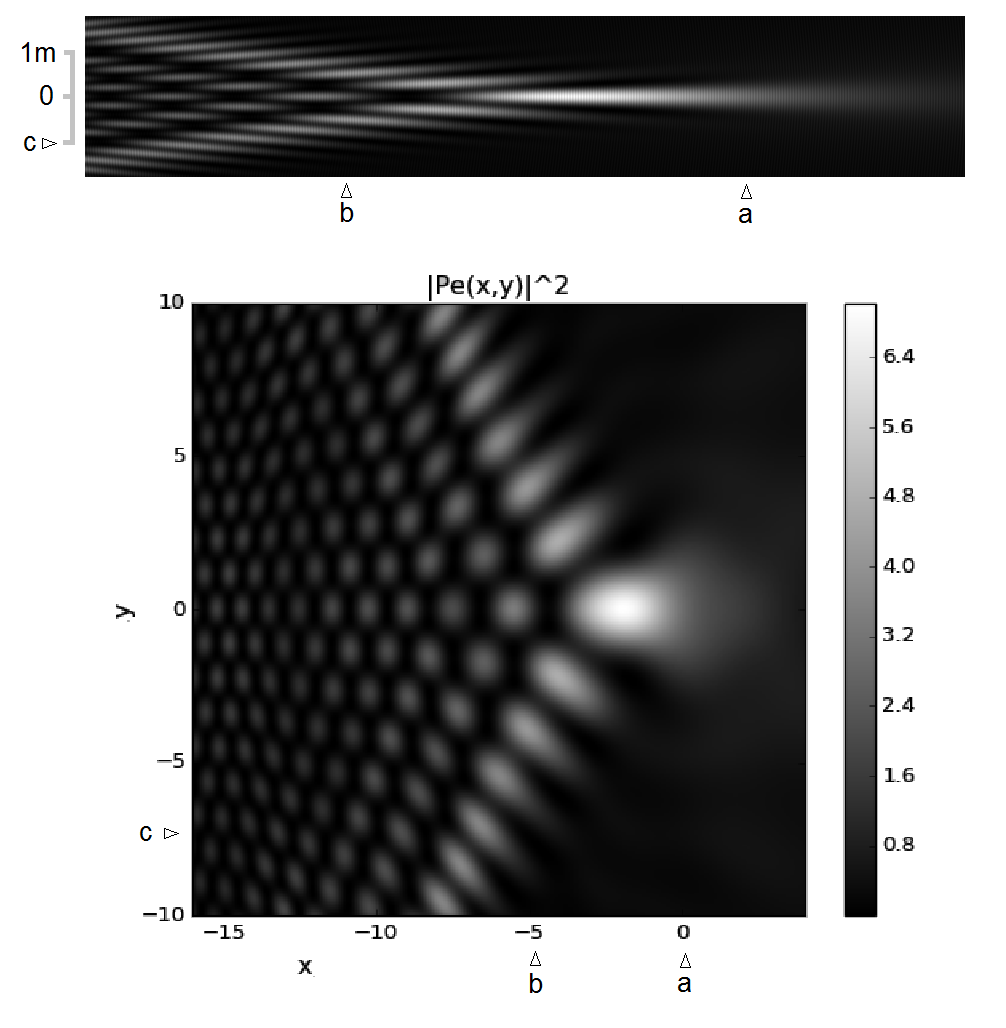}
  \caption{Top: Normalized near-cusp GPSF obtained by numerical integration in (\ref{u}, \ref{Inteps}) for $q\epsilon=3700$. This corresponds, for example, to $\beta=\pi/2$ and $\lambda=10^{-6}$m at 550AU ($d_{\rm astroid}\approx d_{\max}\approx 560$ meters, the aspect ratio is preserved). Bottom: Square of the absolute value of the Pearcey integral.}\label{Pearcey2}
\end{figure}

In difference from the symmetric case, where the circular invariant diffraction pattern has radial oscillations, the near-cusp pattern in the $q\epsilon\gg 1$ case has a complicated two-dimensional lattice-like structure (see Figures \ref{muqe}, \ref{Pearcey2}). The latter transforms towards locally one dimensional structure of smaller intensity (\ref{Folds}) as one moves along the caustic away from the cusp.

\begin{figure}
  \centering
  \includegraphics[width=77mm]{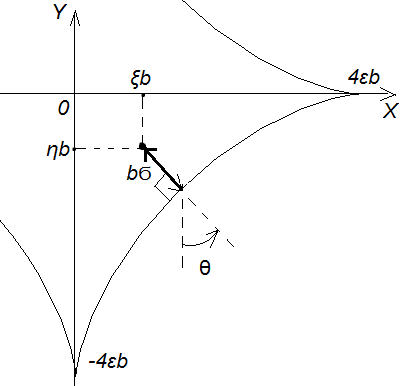}
  \caption{Correspondence between the Cartesian and caustic linked coordinates in the observer plane}\label{Caustic}
\end{figure}

We now evaluate the GPSF in the vicinity of caustic folds (regular points of caustic) far from the cusps\footnote{Description of diffraction pattern in a fold neighborhood is extensively  presented in the literature on the gravitational lensing. For a review, see e.g. \cite{SEF}, \cite{ND}}.

It is convenient to introduce the caustic-linked coordinates $\sigma, \theta$ (see Figure \ref{Caustic})
\begin{equation}
\xi = 4\epsilon\cos^3\theta-\sigma\sin\theta,  \quad \eta = -4\epsilon\sin^3\theta+\sigma\cos\theta
\label{xi_eta}
\end{equation}
with $\sigma$ being the dimensionless length of the perpendicular from the point $(\xi,\eta)$ to the caustic fold. According to (\ref{F0}), in these coordinates
\begin{equation}
F=\frac{1}{2\pi}\int_0^{2\pi} e^{iV}d\phi, \quad V=q\sigma\sin(\theta-\phi)+\epsilon q\left(\cos2\phi+4\sin^3\theta\sin\phi-4\cos^3\theta\cos\phi\right) .
\label{Int}
\end{equation}
Introducing new integration variable $\varphi$ as well as making the change of $\sigma$:
$$
\phi=\theta+\frac{\varphi}{(3q\epsilon\sin 2 \theta)^{1/3}}, \quad \tilde{\sigma} = \frac{-\sigma q}{(3\epsilon q\sin2\theta)^{1/3}} ,
$$
we rewrite (\ref{Int}) as
\begin{equation}
F=\frac{1}{2\pi(3\epsilon q\sin2\theta)^{1/3}}\int_{-(3q\epsilon\sin2\theta)^{1/3}\pi}^{(3q\epsilon\sin2\theta)^{1/3}\pi} e^{iV}d\varphi, \quad V=\tilde{\sigma}\varphi+\frac{1}{3}\varphi^3+\tilde{W} ,
\label{FA}
\end{equation}
where
$$
\tilde{W}=\frac{\varphi^3}{(3\epsilon q\sin 2\theta)^{1/3}}\left[\frac{\tilde{\sigma}\tilde{W}_1+\varphi^2\tilde{W}_2}{(3\epsilon q\sin 2\theta)^{1/3}}+\varphi\tilde{W}_3\cot2\theta\right].
$$
In the above equation $\tilde{W}_1$, $\tilde{W}_2$, $\tilde{W}_3$ are bounded functions ($\max(|\tilde{W}_1|, |\tilde{W}_2|, |\tilde{W}_3|)\le 21/20$) of the single variable $\phi=\frac{\varphi}{(3\epsilon q\sin 2\theta)^{1/3}}$, $\phi=-\pi..\pi$
$$
\tilde{W}_1=\frac{\sin\phi-\phi}{\phi^3}, \quad \tilde{W}_2=2\frac{\sin\phi-\phi+\frac{\phi^3}{6}-2\sin2\phi+4\phi-\frac{(2\phi)^3}{3}}{\phi^5}, \quad
\tilde{W}_3=\frac{\cos2\phi-4\cos\phi+3}{\phi^4}.
$$
When the conditions
\begin{equation}
|\tilde{\sigma}|\ll (\epsilon q\sin 2\theta)^{2/3}, \quad |\sin2\theta| \gg\frac{1}{(\epsilon q)^{1/4}}, \quad q\epsilon \gg 1
\label{Ai}
\end{equation}
hold, the $\tilde{W}$-term can be neglected in (\ref{FA}), since $\epsilon q\gg1$. Therefore
$$
F=\frac{1}{2\pi (3q\epsilon\sin2\theta)^{1/3}}\int_{-\infty}^{\infty} \exp i\left[\tilde{\sigma}\varphi+\frac{\varphi^3}{3}\right]d\varphi = \frac{{\rm Ai}(\tilde{\sigma})}{(3q\epsilon\sin2\theta)^{1/3}}, 
$$
where ${\rm Ai}(\tilde{\sigma})$ is the Airy function. Finally we get
$$
\frac{\mu}{\mu_0}=\frac{1}{(3q\epsilon\sin2\theta)^{2/3}}{\rm Ai}^2\left(\frac{-\sigma q}{(3\epsilon q\sin2\theta)^{1/3}}\right), \quad |\sigma| \ll \epsilon|\sin2\theta|, \quad |\sin 2\theta| \gg\frac{1}{(\epsilon q)^{1/4}} .
$$
In terms of the unscaled distance from the caustic fold ${\cal D}=\sigma b$, $b=\sqrt{2r_gZ}$ (see Figure \ref{Caustic})
\begin{equation}
\mu=\frac{4\pi^2 r_g }{\lambda K^2}{\mathrm{Ai}}^2\left( \frac{ -2\pi {\cal D}}{K\lambda}\sqrt{\frac{2r_g}{Z}}\right), \quad K=\left(\frac{12\pi\epsilon r_g}{\lambda}\sin2\theta\right)^{1/3} .
\label{Folds}
\end{equation}
The validity domain of (\ref{Folds})
\begin{equation}
|{\cal D}|\ll b\epsilon|\sin2\theta|, \quad  |\sin2\theta| \gg\frac{1}{(\epsilon q)^{1/4}}
\label{small_folds}
\end{equation}
is obviously contained in (\ref{R_validity}). The maximum of ${\rm Ai}(\tilde{\sigma})$ is reached at $\tilde{\sigma}=-1.02$, which satisfies (\ref{Ai}). The diffraction pattern in the fold neighborhood, far from the cusps, is locally one dimensional with the oscillation scale $\sim\lambda |3\epsilon q\sin2\theta|^{1/3}/\alpha_E$ depending on the fold coordinate $\theta$. Also for maximal magnification at fixed $\theta$ we have
$$
\frac{\max_{\cal D}\mu({\cal D},\theta)}{\mu_0}=\frac{\max {\rm Ai}^2}{(3\epsilon q\sin2\theta)^{2/3}}\approx \frac{0.29}{(3\epsilon q\sin2\theta)^{2/3}}, \quad \frac{\max_{\cal D}\mu({\cal D},\theta)}{\mu_\epsilon}\approx \frac{0.56}{(\epsilon q)^{1/6}\sin^{2/3}2\theta}  .
$$
Taking into account that (\ref{Folds}) is valid for $|\theta| \gg (\epsilon q)^{-1/4}$ and $\epsilon q\gg 1$, we conclude that
$$
\max_{\theta,\cal D}\mu({\cal D},\theta)<\mu_\epsilon  .
$$
This confirms an obvious fact that in a neighborhood of folds the intensity is smaller than the maximum $\mu_\epsilon$ in the cusp region.

According to (\ref{Pearcey}, \ref{Folds}) the GPSF oscillates and amplitude of oscillations falls as one moves away from the caustic and the ``far field" GPSF can be approximated by (\ref{eS}). \\

{\bf 3.} Concluding this section we mention the case when the observer is on the $z$-axis, i.e. when $X=Y=0$:  Here, the integral (\ref{F0}) is expressed in terms of the zero-order Bessel function and
\begin{equation}
\mu_{\rm axis}=\pi q J_0^2(\epsilon q), \quad \frac{\mu_{\rm axis}}{\mu_0}=J_0^2(\epsilon q),
\label{on_axis}
\end{equation}
which is in agreement with numerical results (see Figure \ref{muqe}).

When $\epsilon q\gg 1$, i.e. when the size of the caustic exceeds greatly the diffraction radius \footnote{i.e. when an on-axis observer sees the perfect ``Einstein cross" image in the lens plane.}
$$
\frac{\mu_{\rm axis}}{\mu_0} = \frac{2}{\epsilon q}\cos^2(\epsilon q-\pi/4), \quad \epsilon q \gg 1
$$
which is in agreement with (\ref{eS}).

\section{PSF in Focal Plane of Telescope}

The GPSF (\ref{u}, \ref{Inteps}) is, in fact, a point spread function for a ``zero-aperture" telescope that can be used only for the intensity scan in the observer plane. One should use the PSF of a compound system of the gravitational lens and a telescope (for more details see e.g. \cite{N}) if the diffraction resolution of a telescope is finer than the angular radius of the Einstein ring $\alpha_E$.

\begin{figure}
  \centering
  \includegraphics[width=155mm]{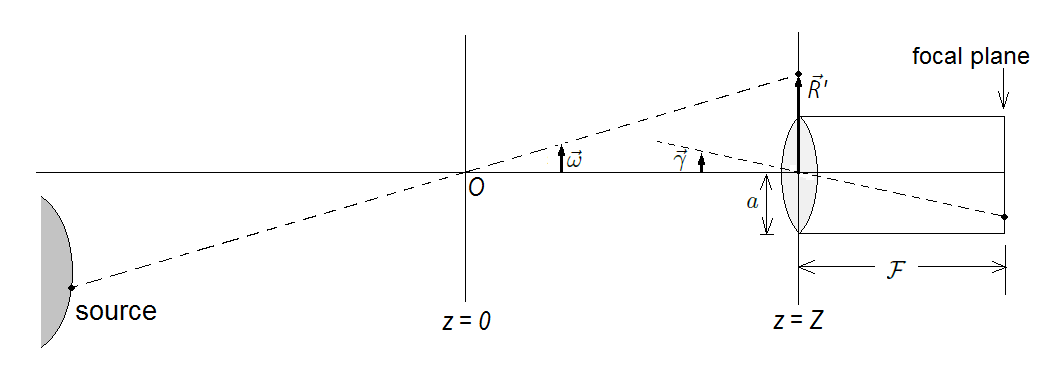}
  \caption{Schematic illustration of the observation angle $\vec{\gamma}$ and the point source angle $\vec{\omega}$. $\vec{R'}=(X',Y')$ is vector of the heliocentric projection of the point source to the observer plane}.
\label{Telescope}
\end{figure}

In the Fraunhofer approximation for the telescope lens of the focal length ${\cal F}$, the PSF in the telescope focal plane is expressed through the Fourier transform of the complex field amplitude at aperture
\begin{equation}
{\cal M}(\vec{\gamma};\vec{\omega})= \left|w(\vec{\gamma};\vec{\omega})\right|^2, \quad
w(\vec{\gamma};\vec{\omega})=\frac{k}{2\pi i{\cal F}}\int_{\vec{R}\in{\mathrm {Aperture}}} u(\vec{R}-\vec{\omega}Z)e^{ik\left(\vec{\omega}+\vec{\gamma}\right)\vec{R}}d^2R ,
\label{muT}
\end{equation}
where $\vec{\gamma}$ and $\vec{\omega}$ are the observation and the point source angles correspondingly (see Figure \ref{Telescope}), and $\vec{R}=(X,Y)$ are coordinates in the aperture (observer) plane. The complex amplitude of the EM field at the aperture plane $u(\vec{R}-\vec{\omega}Z)e^{ik\vec{\omega}\vec{R}}$ is expressed through $u(\vec{R})$ given by (\ref{HuygensFreshnel}). This expression is obtained by application of the small rotation \footnote{Here we neglected ${\cal O}(\omega^2)$ terms, since the variation of the optical path across the aperture due these terms is of order $a\omega^2$, where $a$ is the radius of aperture. Even for $\omega \sim \alpha_E(Z_{\rm min})$ this variation is $\sim 10^{-10}$m, not to mention $\omega \sim$ apparent sizes of objects to be observed.}
$$
\vec{R}\to\vec{R}-\vec{\omega}Z, \quad Z\to Z+ \left(\vec{\omega}\cdot\vec{R}\right)
$$
to the spatial arguments of the complex field amplitude $e^{ikZ}u(\vec{R})$. We recall that the phase factor $e^{ikZ}$ was dropped in (\ref{HuygensFreshnel}), so we restored it to get the complex amplitude at aperture. After getting the amplitude by application of the above rotation, this factor can be dropped again in (\ref{muT}).
For simplicity, we will perform our computations modulo common phase factors.

After substituting (\ref{HuygensFreshnel}) to (\ref{muT}) one gets (up to a common phase factor)
\begin{equation}
w=\frac{k^2}{4\pi^2{\cal F}Z}\int e^{ik\left[
\frac{(\vec{\omega}Z+\vec{r}_\perp)^2}{2Z}-\Psi(\vec{r}_\perp)
\right]}
e^{ik\left(\vec{\gamma}-\frac{\vec{r}_\perp}{Z}\right)\vec{R}+ik\frac{R^2}{2Z}}
{\cal A}(\vec{R})d^2Rd^2r_\perp ,
\label{w_integral}
\end{equation}
where $\vec{r}_\perp=(x,y)$ and ${\cal A}$ is the aperture function, i.e.
$$
{\cal A}(\vec{R})=\left\{\begin{array}{l}
1, \quad \vec{R}\in{\mathrm {aperture}} \\
0, \quad \vec{R}\not\in{\mathrm {aperture}}
\end{array}
\right. .
$$
The last term $ik\frac{R^2}{2Z}$ in the last exponential of eq. (\ref{w_integral}) can be dropped when $a\ll \sqrt{\lambda Z}$, where $a$ is the radius of aperture \footnote{For $Z\sim 1000$AU and $\lambda\sim 10^{-6}$m, $\sqrt{\lambda Z}\sim 10^4$m, so this term can be dropped for any realistic aperture}. Therefore
\begin{equation}
w=\frac{k}{2\pi Z}\int e^{ik\left[
\frac{(\vec{\omega}Z+\vec{r}_\perp)^2}{2Z}-\Psi(\vec{r}_\perp)
\right]}{\cal A}_k\left(\vec{\gamma}-\frac{\vec{r}_\perp}{Z}\right)d^2r_\perp ,
\label{w}
\end{equation}
where function ${\cal A}_k$ is proportional to the Fourier transform of the aperture function
$$
{\cal A}_k(\vec{\vartheta})=\frac{k}{2\pi {\cal F}}\int {\cal A}(\vec{R})e^{ik\vec{\vartheta}\vec{R}}d^2R .
$$
For a circular aperture of radius $a$, ${\cal A}_k$ can be expressed through the Bessel function of the first order
\begin{equation}
{\cal A}_k(\vec{\vartheta})=\frac{a}{|\vec{\vartheta}|{\cal F}}J_1(ka|\vec{\vartheta}|) .
\label{Ak}
\end{equation}
As in the case of the GPSF  \footnote{As in the case of the GPSF, we drop the $cT(Z)$-term in (\ref{Psi}), since the variation of the optical path over the aperture due to this term is at most $2r_ga\omega/Z\approx a\alpha_E^2\omega$. Indeed, even for $\omega\sim \alpha_E$ this variation is $\sim a\times 10^{-15}$, not to mention $\omega \sim$ apparent sizes of object to be observed.}, the stationary phase integration in $\rho$ (recall that $r_\perp=b\rho$, see (\ref{r})) can be performed in (\ref{w}). Indeed, a variation of the argument of the Bessel function in (\ref{w}, \ref{Ak}) across the stationary phase region is $\sim ka\delta r_\perp/Z$, where $\delta r_\perp \sim \sqrt{\lambda Z}$ is the width of this region. Therefore, such a variation can be neglected when $a\ll \sqrt{\lambda Z}$. The last condition holds for any realistic aperture.

\begin{figure}[t]
  \centering
  \includegraphics[width=190mm]{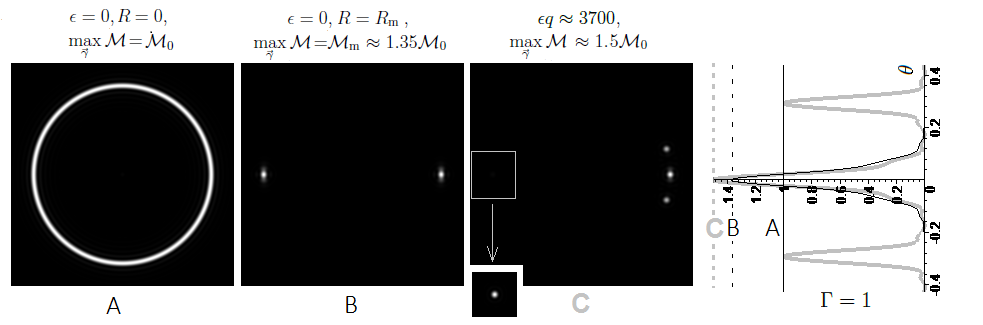}
  \caption{
  Normalized focal plane images for $p:=ka\alpha_E=60$, which would correspond e.g. to 2-meter aperture ($a=1$m) at $\lambda\approx 1\mu$m, and $Z\approx 550$AU. Images A and B show the $\epsilon=0$ case. Fig A: Telescope is centered at the caustic line. Fig B. Maximum of the symmetric case ${\cal M}_{\rm m}:=\max_{\vec{\gamma}, \vec{R}}{\cal M}\vert_{\epsilon=0}$ is observed at $\theta=0$ and $\theta=\pi$. The distance between the observer and the caustic line equals $R_{\rm m}$. Fig C: Non-symmetric case $\epsilon q =3700$ which would correspond to the equatorial PSF at e.g. the above mentioned conditions. The observer coordinates are $X\approx 0.479 \times d_{\rm astroid}$, $Y=0$. The focal plane intensity at $\Gamma=1, \theta=0$ exceeds the maximum ${\cal M}_{\rm m}$ of the degenerate case by the factor $\approx 1.1$. The plots of ${\cal M}/{\cal M}_0$ at $\Gamma=1$ in vicinity of point $\theta=0$ are shown on the right figure. Intensities of the images A and B are plotted by black thin solid lines, while the thick solid gray curve corresponds to the image C. Note that in the case C there are 4 geometrical optics images.  The maxima corresponding to three of four images  (1 global and 2 local) are visible on the Panel C. The local maximum at $\Gamma=1,\theta=\pi$ (corresponding to the fourth geometrical optics image, see Figure \ref{Images}) approximately equals $0.04{\cal M}_0$. It is too weak in comparison with the global maximum ${\cal M}\vert_{\Gamma=1,\theta=0}\approx 1.51{\cal M}_0$ to be noticeable on the normalized gray-scale panel C. This local maximum is shown on the separate sub-panel in the lower left corner of the panel C: For better visibility the brightness inside the selected square is increased by $\approx 33$ times.
  }\label{Rings}
\end{figure}

When condition (\ref{R_validity}) holds, one can integrate over the unit circle $\rho=1$ in the observer plane and
\begin{equation}
{\cal M}=\mu_0\left(\frac{ka^2}{2{\cal F}}\right)^2|F|^2, \quad F=\frac{1}{\pi}\int_0^{2\pi}e^{ik\alpha_E\left(\epsilon b\cos 2\phi-X\cos\phi-Y\sin\phi\right)}h\left(ka\alpha_E\left|e^{i\phi}-\Gamma e^{i\theta}\right|\right)d\phi ,
\label{FT}
\end{equation}
where function of one variable $h(\cdot)$ is defined as
$$
h(x)=J_1(x)/x
$$
and $\Gamma, \theta$ are the dimensionless polar coordinates in the focal plane
$$
\vec{\gamma}=(\alpha_E\Gamma\cos\theta, \alpha_E\Gamma\sin\theta) .
$$
In eq. (\ref{FT}), $X,Y$ denote the deviation of the observer from the heliocentric projection of the point source ($(X,Y)$ equals $-\vec{R'}$ on Figure \ref{Telescope}):
$$
(X,Y)=-\vec{R'} = -\vec{\omega}Z .
$$
If an aperture is much more smaller than the diffraction radius $\lambda/\alpha_E$, the PSF (\ref{FT}) is independent of $\theta$ and it is proportional to the GPSF (the focal plane image is  rather the Airy spot than the limbs for such apertures). On the other hand, when an aperture is big enough, so that the telescope diffraction resolution is much finer than $\alpha_E$ and when the object apparent size $\Delta\omega_{\rm max}$ is much more smaller than the diffraction resolution of telescope (i.e. for $a\alpha_E\gg\lambda$ and $a|\Delta\omega_{\rm max}|\ll\lambda$) this function is ``concentrated" (within the telescope diffraction limit) in the neighborhood of the ``Einstein circle" $\Gamma=1$ (as on e.g. Figures \ref{Rings}, \ref{Points}).

Note that the ratio ${\cal M}/{\cal M}_0$, where ${\cal M}_0=\max_{\vec{\gamma}}{\cal M}\vert_{\vec{R}=0,\epsilon=0}$ and $\vec{R}=(X,Y)$, is essentially a function of six arguments
$$
\frac{\cal M}{\cal M}_0 = g\left(\frac{\vec{\gamma}}{\alpha_E}, \frac{p\vec{R}}{a}, \epsilon q, p \right), \quad p:=ka\alpha_E .
$$
Thus, our results are scalable. For example, $\max_{\vec{\gamma}, \vec{R}}{\cal M}/{\cal M}_0$ for $\lambda=10^{-6}$m, $a=1$m and $Z=550$AU is the same as for $\lambda = 0.5\times 10^{-6}$m, $a\approx 0.71$m and $Z=1100$AU.

It is also important to note that, in difference from the GPSF $\mu$, the maximum of the focal plane PSF ${\cal M}$ is not necessarily smaller for equatorial observations in comparison with polar observations, when the aperture is big enough (see Figure \ref{Rings}).

In more detail: Consider first the $\epsilon=0$ case.\\

\begin{figure}
  \centering
  \includegraphics[width=105mm]{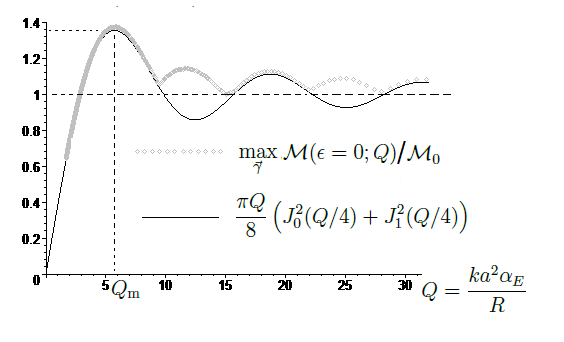}
  \caption{Ratio of the maximal focal plane intensity for $R\not=0$ to that for $R=0$ in the $\epsilon=0$ case is shown by the gray point plot. It is obtained by numerical integration in (\ref{FT}) for  $a\alpha_E/\lambda\approx 10$. The ratio (vertical axis) is plotted against $Q=ka^2\alpha_E/R$ (horizontal axis).
  The ratio given by eq.(\ref{MtoMring}) is plotted by the black line. Note that, at fixed $R$, the number of the focal plane points where maximum is reached is either two or four. The two point maxima (i.e. maxima at $\theta=0$ and $\theta=\pi$) correspond to values of $Q$ at which the above two plots (approximately) coincide. The maximum $\max_{\vec{\gamma}, \vec{R}}{\cal M}\vert_{\epsilon=0}={\cal M}_{\rm m} \approx 1.35{\cal M}_0$ is a two-point maximum. It is seen when $Q=Q_{\rm m}\approx 5.7$ (see Figure \ref{Rings}B).}
\label{Max_Sym}
\end{figure}

{\bf 1)} The numerical integration in (\ref{FT}) shows (see Figure \ref{Max_Sym}) that the focal plane PSF, as a function of $\vec{\gamma}$ and $\vec{R}$ (at fixed $a$, $\lambda$, $Z$ and ${\cal F}$), reaches its global maximum
$$
{\cal M}_{\rm m}:=\max_{\vec{\gamma}, \vec{R}}{\cal M}\vert_{\epsilon=0}
$$
when the telescope is placed at some non-zero distance $R=R_{\rm m}$ from the caustic line. The maximum is reached at $\vec{\gamma}$ corresponding to positions of the geometrical optics images. Let us now find ${\cal M}_{\rm m}$ and $R_{\rm m}$ analytically.

Without loss of generality we can set $X=R\ge 0$, $Y=0$, so that one of the geometrical optics images  \footnote{When $\epsilon=0, R\ll \sqrt{\lambda Z}, a\ll {\sqrt{\lambda Z}}$ and $\lambda\ll r_g$, the PSF is symmetric wrt central inversion $\theta\to\theta+\pi$ . } would be at $\Gamma=1$, $\theta=0$. Considering the case when both the aperture and $R$ are much bigger than the diffraction radius (i.e. $a \gg \lambda/\alpha_E$ and $R \gg \lambda/\alpha_E$) we can approximate value of $F$ in eq. (\ref{FT}) at $\Gamma=1$, $\theta=0$ as follows
\begin{equation}
F\vert_{\epsilon=0,\Gamma=1, \theta=0, Y=0}\approx\frac{1}{\pi}\int_{-\infty}^\infty e^{ik\alpha_E R\phi^2/2}h(ka\alpha_E\phi)d\phi = \frac{1}{\pi ka\alpha_E}\int_{-\infty}^\infty e^{\frac{i\varphi^2}{2Q}}h(\varphi)d\varphi ,
\label{FCmax}
\end{equation}
where
$$
Q=\frac{ka^2\alpha_E}{R} = \frac{pa}{R}.
$$
We would like to compare the intensity at $R\not=0, \theta=0$ with maximal intensity of the circularly symmetric ring seen by the on-axis observer. From (\ref{FT}) it follows that for such a ring (up to a common, $\theta$-dependent phase factor)
$$
F\vert_{\epsilon=0,\Gamma=1, R=0}= \frac{1}{\pi}\int_{-\pi}^\pi h(2ka\alpha_E\sin\frac{\phi}{2})d\phi\approx\frac{2}{\pi ka\alpha_E}, \quad a\gg \lambda/\alpha_E.
$$
Therefore (see (\ref{FT})), the maximal intensity of the circularly symmetric ring seen by the on-axis observer equals
$$
{\cal M}_0:=\max_{\vec{\gamma}} {\cal M}\vert_{\epsilon=0, R=0}=\mu_0\left(\frac{a}{\pi{\cal F}\alpha_E}\right)^2=\frac{q}{\pi} \left(\frac{a}{{\cal F}\alpha_E}\right)^2, \quad a\gg \lambda/\alpha_E .
$$
Note that for small apertures $F\vert_{\epsilon=0,\Gamma=1, R=0}\approx 1$, and therefore in the asymptotic cases we have
$$
{\cal M}_0=\left\{
\begin{array}{ll}
\mu_0\left(\frac{a}{\pi{\cal F}\alpha_E}\right)^2, & a\gg \lambda/\alpha_E .\\
\mu_0\left(\frac{ka^2}{2{\cal F}}\right)^2, & \quad a\ll \lambda/\alpha_E
\end{array}
\right. .
$$
For an intermediate range of apertures one has to evaluate ${\cal M}_0$ numerically. We will take ${\cal M}_0$ as a reference value in what follows.

Integral in eq. (\ref{FCmax}) can be expressed in terms of the Bessel functions \footnote{Here $\int_{-\infty}^\infty e^{\frac{i\varphi^2}{2Q}}\frac{J_1(\varphi)}{\varphi}d\varphi=\frac{\sqrt{2\pi iQ}e^{-iq/4}}{2}\left(J_0(Q/4)+iJ_1(Q/4)\right)$ . } and we get
\begin{equation}
\frac{{\cal M}(\epsilon=0; \Gamma=1, \theta=0, R)}{{\cal M}_0}\approx\frac{\pi Q}{8}\left(J_0^2(Q/4)+J_1^2(Q/4)\right), \quad a\gg \lambda/\alpha_E.
\label{MtoMring}
\end{equation}
This ratio can exceed unity and takes the biggest value $\approx 1.35$ at $Q=Q_{\rm m}\approx 5.7$ (see Figure \ref{Max_Sym} and Figure \ref{Rings}B). In other words, the maximum $\max_{\vec{\gamma}, R}{\cal M}={\cal M}_{\rm m} \approx 1.35{\cal M}_0$ is seen when the telescope is placed at the distance $R=R_{\rm m}$ from the caustic line, where
\begin{equation}
R_{\rm m}= k\alpha_Ea^2/Q_m \approx 1.1\alpha_E a^2/\lambda, \quad a\gg \lambda/\alpha_E .
\label{Rm}
\end{equation}
Note that once an aperture is much bigger than the diffraction radius, the similar condition for the distance $R_{\rm m}\gg \lambda/\alpha_E$ is satisfied automatically.

The domain of validity of eq.(\ref{Rm}) is also restricted by the condition (\ref{R_validity}), i.e. $R_{\rm m}\ll\sqrt{\lambda Z}$ or
$$
\lambda\gg (a^4 \alpha_E^2/Z)^{1/3}=(2r_ga^4/Z^2)^{1/3}.
$$
For $a\sim 1$m and $Z\sim 1000$AU this restriction reads as $\lambda\gg 10^{-8}$m.

The fact that $\max_{\vec{\gamma}, \vec{R}}{\cal M}\vert_{\epsilon=0}$ is not seen when the telescope is centered exactly at the caustic line is explained as follows: Although the telescope at $R=R_{\rm m}$ takes less total energy flux, the size of the bright limb/spot in the focal plane is also smaller (compared to the full ring for telescope at $R=0$), so the maximal flux density through the focal plane happens to be a bit bigger at $R=R_{\rm m}$ than at $R=0$.
At big distances from the caustic line, the characteristic size of the focal plane images is defined only by the diffraction limit of the telescope and the ratio $\max_{\vec{\gamma}}{\cal M}(\epsilon=0; R)/{\cal M}_0$ decays proportionally to $1/R$ (i.e. proportionally to $Q$) when $R\gg R_{\rm m}$ (i.e. $Q\ll Q_{\rm m}$) (see Figure \ref{Max_Sym}).\\

Now, let us return to the non-degenerate $\epsilon\not=0$ case. \\

{\bf 2)} Here, analysis of behaviour of the PSF is more involved, so we present only some general comments and preliminary (mostly numerical) results.

\begin{figure}[t]
  \centering
  \includegraphics[width=175mm]{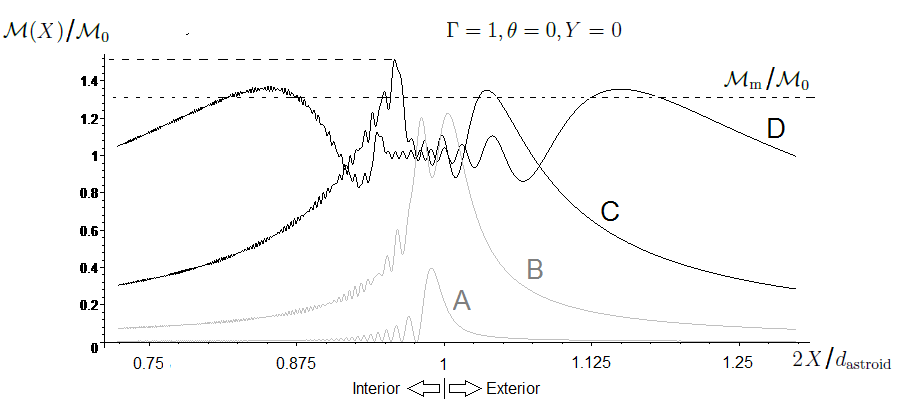}
  \caption{${\cal M}/{\cal M}_0$ is evaluated at the point of the focal plane corresponding to the geometric optic image (at $\Gamma=1$, $\theta=0$) for different positions of observer. The observer moves along the $X$ axis in vicinity of the cusp at $X=d_{\rm astroid}/2, Y=0$. Numerical evaluation of (\ref{FT}) is performed for $\epsilon q\approx 3700$. Values of $p:=ka\alpha_E$ for the curves A,B,C,D are $7.5,30,60,120$ respectively. For equatorial observations at $Z=Z_{\rm min}$ and $\lambda=10^{-6}$m, the curves A, B, C and D would correspond to the apertures of $0.25, 1, 2$ and $4$ meters. In case of curves C and D the maximum of the PSF exceeds that ${\cal M}_{\rm m}$ of the symmetric case: For instance, for curve C: $\max_{X}{\cal M}\vert_{\Gamma=1, \theta=0, Y=0}\approx 1.1 \times {\cal M}_{\rm m}\approx 1.5 \times {\cal M}_0$. The focal plane image corresponding to the maximum of curve C is shown at the Figure \ref{Rings}C. As aperture becomes larger, ${\cal M}$ behaves similarly to the $\epsilon=0$ case: $\max {\cal M}/{\cal M}_0$ decreases as $a$ increases and tends to that of the symmetric case ${\cal M}_{\rm m}/{\cal M}_0$ (c.f. curves C and D). Note the oscillatory behavior of curves in an interior of the astroid (i.e. for $2X/d_{\rm astroid}<1$) due to the diffraction pattern (see Figure \ref{Pearcey2}).}
  \label{mu_aperture}
\end{figure}

\begin{figure}[t]
  \centering
  \includegraphics[width=175mm]{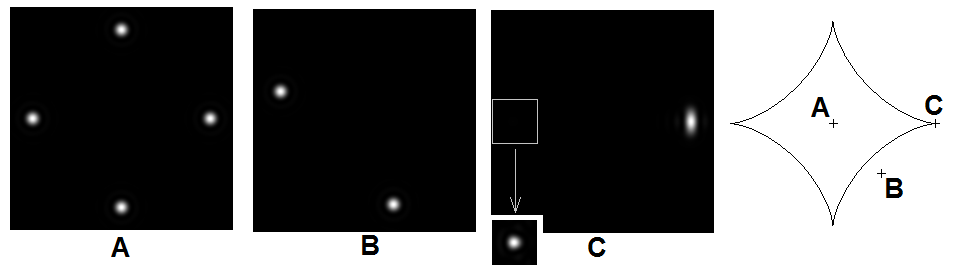}
  \caption{Normalized focal plane images of the equatorial point source viewed through a $1$-meter telescope ($a=0.5$m) obtained by numerical integration in eq. (\ref{FT}). The parameters $\lambda$, $\beta$ and $Z$ are set as in Figure \ref{Pearcey2}.
  The corresponding positions of the telescope wrt caustic are shown on the right figure. In difference from images {\bf A} and {\bf B}, the image {\bf C} is taken at caustic (at the cusp of astroid): The brightest part of the image {\bf C} has an elongated (limb-like) form, while those of images {\bf A} and {\bf B} are round Airy spots of the telescope lens only.
   The maximum of PSF corresponding to the image {\bf A} (``Einstein Cross") is about $2\times 10^{-2}{\cal M}_0$. For image {\bf B} it is $\approx 1.5\times 10^{-2}{\cal M}_0$. For the image {\bf C} the global maximum exceeds ${\cal M}_0$ and is $\approx 1.2\times {\cal M}_0$. Note that value of the local maximum at $\Gamma=1,\theta=\pi$, corresponding to the ``weak image" of the Figure \ref{Images}D is too small in comparison with the global maximum at $\Gamma=1,\theta=0$ to be noticeable on the panel C of the present normalized gray-scale Figure. This local maximum is shown on the separate sub-panel in the lower left corner of the panel C: For better visibility the brightness inside the selected square is increased by $\approx  127$ times.}
   \label{Points}
\end{figure}

Behaviour of the PSF can be easily described in both small and big aperture limits. When the aperture is much smaller than the scale of the diffraction pattern (i.e. when $\alpha_E$ is much more smaller than diffraction limit of telescope) the focal plane PSF is essentially a product of the GPSF and the PSF of the telescope lens. Therefore
$$
\max{\cal M}/\max{\cal M}(\epsilon=0)\to\mu_\epsilon/\mu_0, \quad a\ll\lambda/\alpha_E .
$$
On the other hand when aperture is much more bigger than the diameter of astroid the maxima of the PSF should be approximately the same in the degenerate and non-degenerate cases. Numerical results as well as analysis below show that the maxima can become approximately the same already at apertures that are much more smaller than $d_{\rm astroid}$ (see Figure \ref{mu_aperture}). To get an analytic estimate for the corresponding range of apertures and absolute maximum of ${\cal M}/{\cal M}_0$ (or that of ${\cal M}/{\cal M}_{\rm m}$) we apply the approach similar to that of the symmetric case.

In more detail: Numerical computations show that when the aperture is big enough, $\max_{\vec{\gamma}, \vec{R}}{\cal M}$ is reached at four symmetrically situated points (two at the $X$-axis and two at the $Y$-axis) and at $\vec{\gamma}$ corresponding to a position of a brightest (at given $X$ or $Y$) geometrical optical image. Therefore, we place an observer at the $X$-axis at the point $X=d_{\rm astroid}/2+\tilde R$, $Y=0$, so that the distance from the cusp is much smaller than the diameter of astroid $|\tilde R|\ll d_{\rm astroid}$. Value of ${\cal M}$ at the point $\Gamma=1$, $\theta=0$ of the focal plane (i.e. at the position of the brightest geometrical optical image) can be obtained in a manner similar to that of the symmetric case (see eq. (\ref{FCmax})). Then, up to a common phase factor, we get
\begin{equation}
F\vert_{\Gamma=1, \theta=0, Y=0}\approx \frac{1}{\pi ka\alpha_E}\int_{-\infty}^\infty \exp \left[i\left(\frac{d_{\rm astroid}}{16k^3\alpha_E^3a^4}\varphi^4+\frac{\varphi^2}{2\tilde Q}\right)\right]h(\varphi)d\varphi, \quad \tilde{Q}=\frac{ka^2\alpha_E}{\tilde R} .
\label{FQa}
\end{equation}
When $\tilde{Q} \sim Q_{\rm m}$, the $\varphi^4$-term in the exponent can be neglected if
$$
a\gg a_0 = \left(\frac{d_{\rm astroid}\lambda^3}{\alpha_E^3}\right)^{1/4} .
$$
In this situation one observes maximum which approximately equals ${\cal M}_m$, i.e.
$$
\max{\cal M}/\max{\cal M}(\epsilon=0)\to 1, \quad {\rm when} \quad a\gg a_0 .
$$
Similarly to the degenerate case, the maximum is observed when $\tilde{R}\approx\pm R_{\rm m}$, i.e. at $X\approx d_{\rm astroid}/2\pm R_{\rm m}$, $Y=0$, which is confirmed by the numerical computations. Note that, due to approximation where the $\varphi^4$ term is neglected in (\ref{FQa}), we got eight equal maxima (four inside and four outside the astroid) instead of four: Although for $a\gg a_0$ the difference between the maxima inside and outside the astroid is negligible (see e.g. curve D of Fig \ref{mu_aperture}), only four or them are the global ones.

In the intermediate range of apertures
$$
a \sim a_0, \quad a\gg \lambda/\alpha_E
$$
the analysis of the PSF becomes non-trivial. However, it follows from (\ref{FQa}) that for $a\gg\lambda/\alpha_E$, the ratio $\max_{\tilde{R}}{\cal M}\vert_{\Gamma=1,\theta=0}/{\cal M}_0$ is approximately a function of $a/a_0$ only. Therefore, it is sufficient to perform a set of computations for a fixed $\epsilon q\gg 1$ and different $p$'s (i.e. different apertures) to approximately get the absolute maximum of ${\cal M}/{\cal M}_0$ (i.e. that for all values of parameters) in the strongly non-degenerate case.

Numerical computations performed for the strongly non-degenerate case show that, once the aperture exceeds some value $\sim a_0$, the maximum of ``non-symmetric" PSF becomes even bigger than that of the degenerate case, i.e. the maximum of ${\cal M}$ exceeds ${\cal M}_{\rm m}$ when the aperture is big enough. However, the difference in these maxima cannot exceed few ($\sim 10$) percents. The results of numerical computations are shown on Figure \ref{mu_aperture}. Examples of the corresponding focal plane images are shown on Figure \ref{Rings}.

Note that $a_0\sim 1$m for equatorial observations at $Z\sim 1000$AU and $\lambda \sim 10^{-6}$m. Therefore, for a telescope of a modest aperture ($\sim 1$m), the maximum of the PSF (PSF as a function of $\vec{\gamma}$ and $\vec{R}$) changes only by at most $\approx 10$ percents when the source is moved from the equatorial plane to the polar axis. \\

Finally, let us estimate the PSF when a telescope whose diffraction resolution is much finer than $\alpha_E$ is placed far away from the caustic.
Here our computations facilitate due to the fact that one can perform the two-dimensional stationary phase integration in (\ref{w}). Then, similarly to the case of the GPSF (cf. (\ref{eS})) , one gets
$$
w( \vec{\gamma},\vec{R})=\sum_j \sqrt{|\mu_j|} e^{i(kS_j-\pi n_j/2)} {\cal A}_k\left(\vec{\gamma}-\frac{\vec{r}_j}{Z}\right),
$$
where $\vec{r}_j=(x_j, y_j)$ are coordinates of the $j$th image in the lens plane and $S_j=S(\vec{R};\vec{r}_j)$, $\mu_j$, $n_j$ are same as in eq. (\ref{eS}). Computation of the PSF then reduces to the summation
$$
{\cal M}=\sum_j|\mu_j| {\cal P}\left(\vec{\gamma}-\vec{r}_j/Z\right)+\sum_{j\not=l}\sqrt{|\mu_j\mu_l|}e^{i(k(S_j-S_l)-\pi (n_j-n_l)/2)} {\cal A}_k\left(\vec{\gamma}-\vec{r}_j/Z\right) {\cal A}_k^*\left(\vec{\gamma}-\vec{r}_l/Z\right),
$$
where ${\cal P(\vec{\vartheta})}=|{\cal A}_k(\vec{\vartheta})|^2$ is the PSF of the telescope lens.

When the angular separation between the geometrical optics images of a point source is much more bigger than the diffraction limit of a telescope, the double sum in the last equation can be dropped. Then the focal plane PSF equals the following convolution
\begin{equation}
{\cal M}(\vec{\gamma};\vec{R})=\int \tilde{\mu}(\vec{\gamma'}){\cal P}(\vec{\gamma}-\vec{\gamma'}) d^2\gamma', \quad \tilde{\mu}(\vec{\gamma'})=\sum_j|\mu_j|\delta(\vec{\gamma'}-\vec{r_j}/Z).
\label{geometrical_psf}
\end{equation}
In other words, for a telescope of reasonably big aperture ($a\alpha_E\gg\lambda$) placed away from caustic, the gravitational lensing can be described by the geometrical optics, while the wave effects due to finite aperture have to be taken into account. Related examples of the focal plane images are shown on Figure \ref{Points}.


\section{Discussion and Conclusions}

In this section we would like to discuss implication of effects related to the quadrupole moment of the sun on prospective observations. For this we first summarize the main results of the present work:

The transversal size of the astroid caustic (due to the quadrupole moment of the sun) could reach several hundred meters at the distances ranging from that of the closest observation (550 AU) and up to several thousand of AU. This size is comparable with sizes of heliocentric projections of possible objects of observation, which are about several kilometers across. At such distances, the diffraction pattern of a monochromatic point source transforms significantly (in the region of interest, at e.g. sub-micrometer wavelengths) when the direction of observation is changed from the one along the sun polar axis to that in the sun equatorial plane. The maximum of the gravitational point spread function (GPSF) can differ up to about two-three orders of magnitude. In the strongly non-degenerate case the maximum of GPSF is reached in a neighborhood of the cusp of a caustic. The GPSF can be expressed in terms of the Pearcey integral in this neighborhood.

On the other hand, behaviour of the PSF of a compound system of the gravitational lens and a telescope depends on the telescope's aperture. If the aperture is small, the focal plane PSF and GPSF are essentially the same. For big apertures (e.g. 2m aperture) the absolute maximum of the focal plane PSF can be even bigger in the non-symmetric case. Although these maxima do not differ very much, the formation of images can be significantly different in the symmetric and non-symmetric case. For instance, in contrast to the symmetric case, in the strongly non-degenerate case an image of a point source never forms a ``bright" ring, but rather consists of small limbs/spots. The focal plane image is not generally centrally symmetric (which can be an advantage \footnote{Private communication with S. Turyshev.}), number of images of a point source can be different etc.

We recall that the GPSF/focal plane PSF are magnifications of a monochromatic point source and not those of a realistic extended object. Therefore, in prospective missions, one does not expect to directly observe diffraction patterns.
However, one needs the PSF for de-convolution of realistic images.

In more detail: the energy flux at the point $\vec{R}=(X,Y)$ of the observer plane radiated by an extended, totally (spatially and temporally) incoherent source, equals the convolution of GPSF and the surface brightness of source
\begin{equation}
I(\vec{R})=\int I_{\rm s}(\vec{R'};q)\mu(\vec{R}-\vec{R'};q)d^2R'dq .
\label{Intensity}
\end{equation}
Here $I_{\rm s}(\vec{R'};q)dq$ is a non-magnified energy flux (times the filtering function) in the interval $q,q+dq$ of the spectrum radiated by a surface element on the source, and $\vec{R'}=(X', Y')$ stands for coordinates of the heliocentric projection of this element to the observer plane (see Figure \ref{Telescope}).

The role of caustic in de-convolution process can be already seen in the geometrical optics: Suppose that one got the intensity (\ref{Intensity}) in the non-symmetric case and then performed de-convolution of $I(\vec{R})$ as if $\mu$ were magnification of the monopole. The finest spatial resolution of such a de-convolution will be of order of the astroid diameter $d_{\rm astroid}$, since both non-symmetric and monopole $\mu(\vec{R})$ have the same asymptotic behavior at $R\gg d_{\rm astroid}$ and they start to differ significantly at $R\sim d_{\rm astroid}$. As has been mentioned before, the typical size of the heliocentric projection of an exo-planet to the observer plane is about several kilometers at $1000$AU and $d_{\rm astroid}$ is about 10 percents of that size. Therefore, the area of a minimal pixel will be of order of $1$ percent of that of the whole de-convoluted image, i.e. changes of $\mu$ play an important role starting from about hecto-pixel level of imaging, not to mention the mega-pixel imaging currently discussed in the literature \cite{T}, \cite{TT}. At the latter level of resolution one has to take the diffraction pattern of PSF into account.

Similar arguments can be applied to the de-convolution of the focal plane images. It is most likely that this type of de-convolution, rather than that of intensity scan (\ref{Intensity}) will be used in a possible mission where a sequence of images of the Einstein rings along a path of a spacecraft across the ``high intensity" region will be taken. The intensity of a ring point in this set, in a sense, encodes ``projection data" taken along a ``section" of the source surface. Therefore, a kind of ``tomographic" reconstruction algorithms should be developed for the de-convolution related with the focal plane PSF.


One might consider a hypothetic possibility of obtaining a relatively high resolution image during a single arbitrary passage in a neighborhood or/and through the heliocentric projection  of an exo-planet. Indeed, at $Z\sim 1000 - 2000$AU, a modest $\sim 1$ meter telescope equipped with coronagraphs resolves the Einstein ring with about $\sim 20$ circumferential elements. Therefore, taking about $\sim 10^2$ samples of rings along e.g. a straight path crossing a high-intensity region could, in principle, result in a kilo-pixel image. Note that at this level of resolution the size and structure of caustic plays an important role.


Completing this section, we would like to mention the influence of the higher multi-pole moments of the sun on the PSF.
Computation of the PSF accounting for higher moments is a straightforward generalisation of the case considered in this work: one should add the higher-order harmonic terms to the potential $\psi$ in (\ref{Psi})
$$
\psi=\mathrm{Re}\left[\log(\zeta)-\sum_{i=2}^{\infty}\epsilon_{n} \zeta^{-n}\right], \quad \zeta=\rho e^{i\phi} ,
$$
where $\epsilon_n$ are complex harmonic moments. Then one could perform the geometric optics analysis and the stationary phase integration with an account of a new potential.

Note, that the corrections to the PSF due to the fluctuations of plasma density
in the solar atmosphere might be even more important than corrections accounting for higher harmonic moments.

\vspace{5mm}

{\large \bf Acknowledgements}

\vspace{5mm}

The author is grateful to A.S. Zhedanov for his suggestion to consider effects of the sun oblateness and to O.Yermolayeva and V.Spiridonov for their help.

\section{Appendix 1}

Below we perform the double check of our main results for the strongly non degenerate case $q\epsilon\gg 1$ ($\epsilon\gg\lambda/r_g$) using algorithms from the geometrical theory of diffraction.

\begin{figure}
  \centering
  \includegraphics[width=157mm]{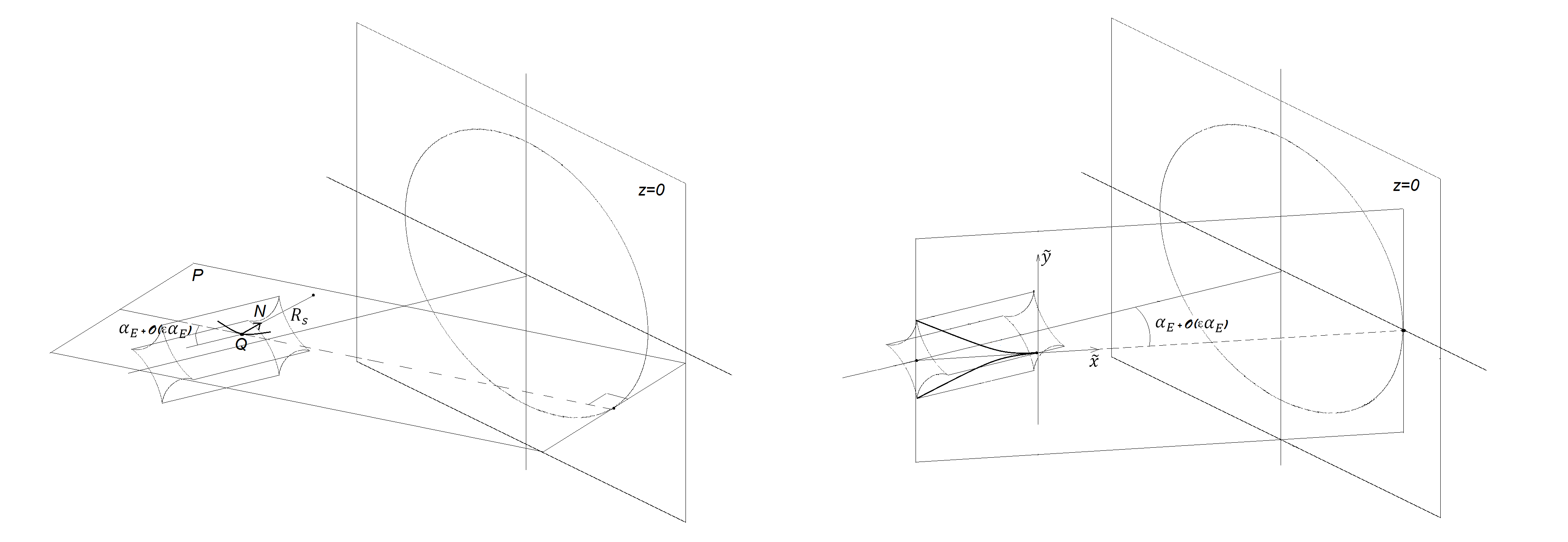}
  \caption{Left: Section of the caustic at fold ($N$ is vector normal to the caustic at $Q$). Right: Section of the caustic at cusp.}\label{Sections}
\end{figure}

We start with the computation of the intensity near a regular point of the caustic. According to the theory of the uniform (caustic) expansions, the intensity magnification in vicinity of a regular point $Q$ of the caustic surface equals to (see e.g. \cite{LL}, \cite{BK})
\begin{equation}
\mu=2\pi U\left(\frac{2k^2}{R_s}\right)^{1/6}{\mathrm {Ai}}^2
\left(-{\cal D}\sqrt[3]{\frac{2k^2}{R_s}}\right) ,
\label{Airy}
\end{equation}
where ${\cal D}$ is the distance to the caustic from its convex side. Here, $R_s$ stands for the radius of curvature of section of the caustic surface by the plane $P$ containing a light ray that is tangent to caustic at $Q$. The plane $P$ also contains the vector normal to the caustic at $Q$ (see Figure \ref{Sections}). The pre-factor $U=U(Q)$ in (\ref{Airy}) is determined by matching the geometrical optics value of magnification (\ref{mu}) in vicinity of the caustic (taking into account the multiplicity of images) with the following asymptotics of (\ref{Airy}) at ${\cal D}\gg (R_s/k^2)^{1/3}$
\begin{equation}
\mu \to \frac{U}{\sqrt{\cal D}} .
\label{mudiff}
\end{equation}
To find all the above values, we use the expansion in the proximity of the critical line (\ref{critical})
\begin{equation}
\rho=\rho_c(\theta)+\Delta \rho .
\label{deltarho}
\end{equation}
It is now convenient to introduce another set of the caustic-linked coordinates $\Delta\rho, \theta$ (also see eq. (\ref{caustic}))
\begin{equation}
(X,Y)=(b\xi_c(\theta), b\eta_c(\theta))+\Delta \vec{r}, \quad \Delta \vec{r} = (2b\Delta \rho \cos\theta, 2b\Delta \rho \sin\theta) .
\label{delta_r}
\end{equation}
Vector $\Delta \vec{r}$ is tangent to the caustic at $Q$. Therefore, for small $\Delta\vec{r}$, the distance from the caustic to the point $(X,Y,Z)$ equals
$$
{\cal D}=\frac{\Delta \vec{r}^2}{2 R_a} = \frac{2b^2}{R_a}\Delta \rho^2 ,
$$
where $R_a$ is the radius of curvature of the astroid (\ref{caustic})
$$
R_a = 6b\epsilon \sin(2\theta) .
$$
The point $(X,Y)$ has two ("strong") pre-images in the close vicinity of the critical line \footnote{and up to four pre-images in total, depending on the observer position relatively to the caustic.} with
$$
\Delta \rho = \pm\frac{1}{2b}\sqrt{2R_a{\cal D}}.
$$
On the other hand, it follows from (\ref{deltarho}) and (\ref{mu}) that away from the caustic $ \mu = \frac{1}{4\Delta \rho} $ and with account of image multiplicity and signs we get
$$
\mu = \frac{1}{2|\Delta \rho|} = \frac{b}{\sqrt{2R_a{\cal D}}} .
$$
Comparing the above equation with (\ref{mudiff}) we get
$$
U=b/\sqrt{2R_a}=b/\sqrt{12\epsilon b\sin 2\theta} .
$$
Also
$$
R_s = R_a/\alpha_E^2=\frac{6\epsilon b\sin(2\theta)}{\alpha_E^2},
$$
since the plane $P$ intersects the $z$-axis under the angle $\alpha_E+{\cal O}(\alpha_E\epsilon)$ (see Figure \ref{Sections}).

Plugging the above values of $U$ and $R_s$ into (\ref{Airy}), with the help of (\ref{parameters1}, \ref{parameters2}), we get the expected final expression (\ref{Folds}) for the near-fold GPSF.

Let us now consider the pattern in regions near the turning points $\theta=0, \pi/2, \pi, 3\pi/2$, where the intensity reaches its maximum. Without loss of generality we take the cusp at $\theta=0$.

According to the theory of the uniform caustic expansions (see eg \cite{BK}, \cite{P}), in the local coordinates of the section plane $\tilde{x}, \tilde{y} $ (see Figure \ref{Sections}), where equation of the caustic has the approximate form
\begin{equation}
\tilde{x}^3=-\frac{9}{8}a\tilde{y}^2,
\label{localcaustic}
\end{equation}
the magnification equals
\begin{equation}
\mu(\tilde{x}, \tilde{y})=W\:\left|\mathrm{Pe}\left(\tilde{x}\left(\frac{6k}{a}\right)^{1/2},\tilde{y}\left(\frac{24k^3}{a}\right)^{1/4}\right)\right|^2 .
\label{muPercey}
\end{equation}
Here, the pre-factor $W$ is determined by matching the geometrical optics value of magnification (\ref{mu}) in the cusp neighborhood
with the corresponding asymptotics of (\ref{muPercey}). It is convenient to set $\tilde{y}=0$ and use the asymptotics of (\ref{muPercey}) for $\tilde{x}\gg\sqrt{\frac{a}{6k}}$
\begin{equation}
\mu(\tilde{x}, 0)\to\frac{\pi W}{\tilde{x}}\sqrt{\frac{a}{6k}} .
\label{mu_W}
\end{equation}
The section plane is parallel to the $y$-axis and intersects the $z$-axis under the angle $\alpha_E+{\cal O}(\alpha_E\epsilon)$ (see Figure \ref{Sections}). Therefore,
\begin{equation}
{\tilde X}=\alpha_E \tilde {x}, \quad {\tilde Y} = \tilde {y} ,
\label{dXY}
\end{equation}
where $({\tilde X}, \tilde{Y})$ is the deviation from the cusp in the observer $z=Z$ plane. Taking (\ref{dXY}) into account, from (\ref{localcaustic}) and (\ref{caustic}) we get
$$
a=12\epsilon b/\alpha_E^3 .
$$
From (\ref{mu}) it follows that for ${\tilde Y}=0$ and ${\tilde X}>0$, near the cusp $\mu\to \frac{b}{2{\tilde X}}$. Then with the help of  (\ref{mu_W}, \ref{dXY}) we obtain
$$
W=\frac{1}{4\pi}\sqrt{\frac{2kb\alpha_E}{\epsilon}} .
$$
Substituting the above values in (\ref{muPercey}) and taking (\ref{dXY}) into account we get the expected expression (\ref{Pearcey}) for the near-cusp GPSF.

\section{Appendix 2}

The solar atmosphere introduces corrections to the gravitational deflection picture. In the first approximation, the correction $\vec{\alpha}_{\rm pl}$ to the total deflection angle $\vec{\alpha}\to \vec{\alpha}+\vec{\alpha}_{\rm pl}$ due to refraction in the solar plasma is the cylindrically symmetric vector field (for a review see e.g. \cite{TA} and references therein)
$$
\alpha_{\rm pl}=\left(\frac{\lambda}{\Lambda}\right)^2\left[A\left(\frac{R_0}{r_\perp}\right)^{2}+B\left(\frac{R_0}{r_\perp}\right)^{6}+C\left(\frac{R_0}{r_\perp}\right)^{16}\right],
$$
where $\Lambda \approx 50$ meters, $A\approx 1.1$, $B\approx 2.28\times 10^2$, $C\approx 2.952 \times 10^3$. The atmosphere bends the rays
outwards while the gravity bends the rays inwards.  The above correction corresponds to the circularly symmetric contribution $\psi_{\rm pl}=\psi_{\rm pl}(\rho)$ to the dimensionless potential $\psi \to \psi+\psi_{\rm pl}$ (for definitions of $\psi$ and $\rho$ see eqs. (\ref{Psi},\ref{r}))
$$
\psi_{\rm pl}=\epsilon_{\rm pl}\left[A\frac{R_0}{b}\frac{1}{\rho}+\frac{B}{5}\left(\frac{R_0}{b}\right)^{5}\frac{1}{\rho^5}+\frac{C}{15}\left(\frac{R_0}{b}\right)^{15}\frac{1}{\rho^{15}}\right],
$$
where
$$
\epsilon_{\rm pl}=\left(\frac{\lambda}{\Lambda}\right)^2\frac{R_0}{r_g}
$$
and $b=b(Z)$ is given by (\ref{parameters1}).

The effect of the solar plasma is small for the sub-micrometer wavelengths. Indeed, for $\lambda=10^{-6}$m, $\epsilon_{\rm pl}\approx 10^{-10}$.

It is easy to see that due to the circular symmetry of $\psi_{\rm pl}$ and smallness of $\epsilon_{\rm pl}$ the atmospheric refraction can be taken into account in this approximation by the formal re-normalization of the gravitational radius $r_g\to \tilde r_g(Z)$ in all our previous results
$$
\tilde r_g(Z)\approx r_g\left(1-\frac{\alpha_{\rm pl}(Z)}{\alpha_E(Z)}\right),
$$
where $\alpha_{\rm pl}(Z)$ is the $\alpha_{\rm pl}$ evaluated at $r_\perp = b(Z)$. Note that the correction factor $\alpha_{\rm pl}/\alpha_E\approx 5\times 10^{-9}$ for $\lambda = 10^{-6}$m and $Z\approx 1000$AU, i.e. effect of refraction in sub-micrometer range of the EM spectrum is extremely small in this approximation.

Higher order (non-symmetric) corrections to $\psi_{\rm pl}$ are proportional to the product of $\epsilon_{\rm pl}$ and small deformation parameters such as the oblateness coefficient of the columnar density of the solar atmosphere etc.
In principle, the contribution to $\psi_{\rm pl}$ corresponding to this oblateness can be accounted for through an effective ($Z$-dependent) correction of the quadrupole moment of the sun.
This and higher moment corrections can be also discarded in the context of the present work due to their smallness for wavelengths of interest. However, the question of deformations of caustic/PSF due to fluctuations in the solar atmosphere is worth studying in the context of the high-resolution or/and longer wavelength imaging.

\end{document}